\documentclass[twocolumn]{autart}   
\usepackage{graphicx}
\usepackage{natbib}
\usepackage[caption=false]{subfig}
\usepackage{epsfig}
\usepackage{epstopdf}
\usepackage{amsmath}
\usepackage{amssymb}
\usepackage{color}

\DeclareMathOperator*{\argmin}{arg\,min}
\newcommand\aseq{\mathrel{\overset{\makebox[0pt]{\mbox{\normalfont\small a.s.}}}{=}}}

\usepackage{comment}

\begin{document}

\begin{frontmatter}

\title{Chance-Constrained Controller State and \\ Reference Governor\thanksref{footnoteinfo}} 
\vspace{-2mm}

\thanks[footnoteinfo]{This paper was not presented at any IFAC
meeting. Corresponding author N.~Li. Email {\it nanli@umich.edu}}

\author[UM]{Nan Li}\ead{nanli@umich.edu},
\author[UM]{Anouck Girard}\ead{anouck@umich.edu},
\author[UM]{Ilya Kolmanovsky}\ead{ilya@umich.edu}

\address[UM]{Department of Aerospace Engineering, University of Michigan, Ann Arbor, Michigan, USA}
\vspace{-2mm}

\begin{abstract}
The controller state and reference governor (CSRG) is an add-on scheme for nominal closed-loop systems with dynamic controllers which supervises the controller internal state and the reference input to the closed-loop system to enforce pointwise-in-time constraints. By admitting both controller state and reference modifications, the CSRG can achieve an enlarged constrained domain of attraction compared to conventional reference governor schemes where only reference modification is permitted. This paper studies the CSRG for systems subject to stochastic disturbances and chance constraints. We describe the CSRG algorithm in such a stochastic setting and analyze its theoretical properties, including chance-constraint enforcement, finite-time reference convergence, and closed-loop stability. We also present examples illustrating the application of CSRG to constrained aircraft flight control.

\vspace{-2mm}
\end{abstract}


\end{frontmatter}

\section{Introduction}\label{sec:1}

Reference governors (RGs) are add-on control schemes used to protect pre-stabilized, closed-loop systems against violations of pointwise-in-time constraints \citep{garone2017reference}. They do so by monitoring, and modifying when necessary, the reference input to the closed-loop system. Instead of (re-)designing a controller that simultaneously achieves stabilization and constraint enforcement (as well as other performance requirements) as in the model predictive control framework \citep{mayne2000constrained}, the RG setting preserves the existing/legacy architecture of the closed-loop system, while augmenting the nominal system with the ability to handle constraints. 

\begin{figure}[ht!]
\begin{center}
\begin{picture}(400.0, 75.0)
\put(  0,  -2){\epsfig{file=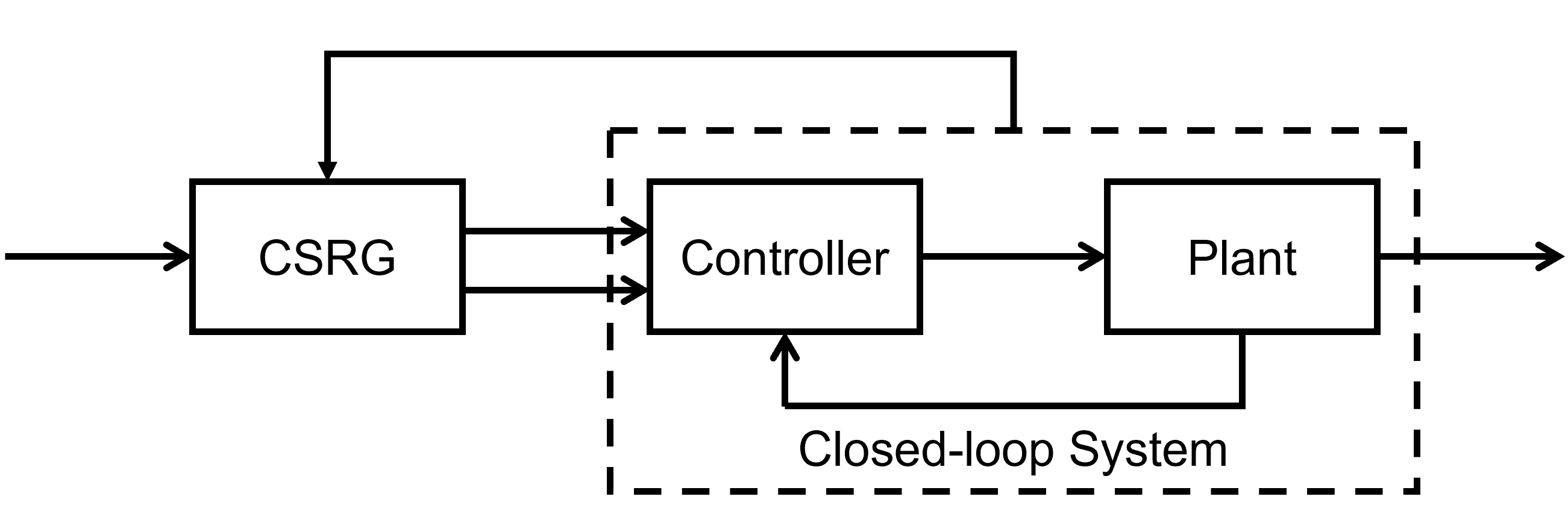,height=1.05in}}  
\small
\put(  6,  40.5){$r(t)$}
\put(  73.5,  44){$v(t)$}
\put(  72,  23){$x_u(t)$}
\put(  142,  40.5){$u(t)$}
\put(  141.5,  18.5){$x_p(t)$}
\put(  211,  40.5){$y(t) \!\in\! \mathcal{Y}$}
\put(  90,  70){$\bar{x}(t)$}
\normalsize
\end{picture}
\end{center}
      \caption{Schematic of the controller state and reference governor (CSRG).}
      \label{fig:CSRG}
\end{figure}
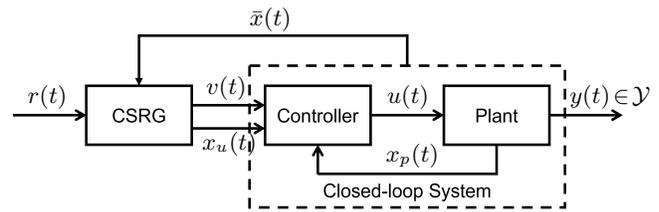

An extension of the RG, called the {\it controller state and reference governor (CSRG)}, was proposed for closed-loop systems with dynamic controllers in \cite{mcdonough2015controller}. The CSRG monitors and modifies not only the reference input to the closed-loop system, but also the internal state of the dynamic controller (see Fig.~\ref{fig:CSRG}). By admitting both controller state and reference modifications, the constrained domain of attraction, i.e., the set of initial states which can be recovered without constraint violation, is enlarged. It is worth noting that the approach of modifying/resetting controller state has also been exploited in classical nonlinear control, mainly for improving control performance. For instance, \cite{bupp2000resetting} proposed a control strategy called the resetting virtual absorber, where the controller state is periodically reset to dissipate energy from a vibrating system. In contrast, CSRG modifies the controller state for enforcing constraints, which has not been broadly investigated before.

Conventional RG schemes, including the CSRG studied in \cite{mcdonough2015controller}, are able to enforce constraints for deterministic systems \citep{bemporad1997nonlinear,bemporad1998reference,borrelli2009reference}, or robustly enforce constraints for systems subject to disturbance inputs that take values in compact sets \citep{gilbert1999fast,casavola2000robust,gilbert2002nonlinear,gilbert2011constrained,li2020log}. However, in many application scenarios, the system is acted on by disturbances which are represented by stochastic models. In this setting, a typical approach is to impose constraints as {\it chance constraints}, i.e., seek probabilistic guarantees on constraint satisfaction \citep{birge2011introduction}.


In recent years, there has been a growing interest in developing control techniques for systems subject to stochastic disturbances and chance constraints, mainly within the stochastic model predictive control framework (see \cite{mesbah2016stochastic,mesbah2019stochastic} and references therein). Latterly, an RG for chance-constrained systems was developed in \cite{kalabic2019reference}, which exhibited theoretical properties that were analogous to the ones for conventional RGs, including chance-constraint enforcement and reference convergence guarantees.

In this paper, we study the CSRG in a stochastic setting. The contributions of this paper are as follows:

\begin{enumerate}
    \item We develop a CSRG scheme for discrete-time linear systems subject to stochastic disturbances and chance constraints. This CSRG scheme operates based on a finitely-determined approximation to a chance-constrained maximal output admissible set and online optimization. In particular, two online algorithms with distinct features are presented.
    \item We analyze theoretical properties of the proposed CSRG scheme, including finite-determinability of the used output admissible set, closed-loop chance-constraint satisfaction, almost-sure finite-time convergence of the modified reference to constant, steady-state constraint-admissible commands, and mean-square stability of the commanded state set-point. These theoretical properties distinguish our CSRG scheme versus several other control techniques for systems subject to stochastic disturbances and chance constraints. For instance, in stochastic model predictive control, guarantees on closed-loop chance-constraint satisfaction and closed-loop stability are typically not easy to establish \citep{mesbah2016stochastic}, especially in the presence of disturbances with unbounded supports (such as the Gaussian disturbance treated in this paper).
    \item We extend the proposed CSRG scheme from its nominal formulation for the case of individual chance constraints to the one that can address joint chance constraints. We describe two approaches to this extension and compare their relative conservativeness, resulting in guidelines for when one approach is preferable over the other.
    \item We illustrate the proposed chance-constrained CSRG scheme through examples representing its application to constrained aircraft flight control.
\end{enumerate}

The paper is organized as follows. In Section~\ref{sec:2}, we define the system and the constraints to be treated. In Section~\ref{sec:3}, we present the chance-constrained CSRG scheme, including the construction of the maximal output admissible set and two online algorithms. In Section~\ref{sec:4}, we analyze theoretical properties of the proposed CSRG scheme, and also extend the scheme from treating individual chance constraints to treating joint chance constraints. In Section~\ref{sec:5}, we present examples illustrating the application of CSRG to constrained aircraft flight control. Finally, Section~\ref{sec:6} concludes the paper.

\section{Problem Statement}\label{sec:2}

In this paper, we consider systems which can be represented by a discrete-time linear model of the form,
\begin{subequations}\label{equ:system_plant}
\begin{align}
x_p(t+1) &= A x_p(t) + B_u u(t) + B_w w(t), \\
y(t) &= C x_p(t) + D_u u(t) + D_w w(t),
\end{align}
\end{subequations}
where $x_p(t) \in \mathbb{R}^{n_{xp}}$ represents the plant state at the discrete-time instant $t \in \mathbb{Z}_{\ge 0}$, $u(t) \in \mathbb{R}^{n_u}$ denotes the control input,
$w(t) \in \mathbb{R}^{n_w}$ denotes an unmeasured disturbance input, and $y(t) \in \mathbb{R}^{n_y}$ represents the system output.

We assume that the following dynamic controller has been designed to stabilize the system \eqref{equ:system_plant},
\begin{subequations}\label{equ:system_controller}
\begin{align}
u(t) &= K_p x_p(t) + K_u x_u(t) + B_v v(t), \\
x_u(t+1) &= A_p x_p(t) + A_u x_u(t) + D_v v(t),
\end{align}
\end{subequations}
where $x_u(t) \in \mathbb{R}^{n_{xu}}$ denotes the controller state, and $v(t) \in \mathbb{R}^{n_v}$ is a reference input determining the set-point of the system. 

The closed-loop system combining \eqref{equ:system_plant} and \eqref{equ:system_controller} can be written in the following compact form,
\begin{subequations}\label{equ:system_cpt}
\begin{align}
\bar{x}(t+1) &= \bar{A} \bar{x}(t) + \bar{B}_v v(t) + \bar{B}_w w(t), \label{equ:system_cpt_1} \\
y(t) &=  \bar{C} \bar{x}(t) + \bar{D}_v v(t) + \bar{D}_w w(t),
\end{align}
\end{subequations}
where $\bar{x}(t) = [x_p^{\top}(t), x_u^{\top}(t)]^{\top}$, and
\begin{subequations}\label{equ:system_cpt_2}
\begin{align}
\bar{A} &= \begin{bmatrix}
A+B_u K_p & B_u K_u \\ A_p & A_u
\end{bmatrix},\, \bar{B}_v = \begin{bmatrix}
B_u B_v \\ D_v
\end{bmatrix},\, \bar{B}_w = \begin{bmatrix}
B_w \\ 0
\end{bmatrix}, \\
\bar{C} &= \begin{bmatrix}
C + D_u K_p & D_u K_u
\end{bmatrix},\, \bar{D}_v = D_u B_v,\, \bar{D}_w = D_w.
\end{align}
\end{subequations}
We make the following assumptions:

{\it Assumption 1:} The matrix $\bar{A}$ is Schur, i.e., all eigenvalues of $\bar{A}$ are strictly inside the unit disc.

Since $\bar{A}$ corresponds to the closed-loop system consisting of the plant \eqref{equ:system_plant} and the stabilizing controller \eqref{equ:system_controller}, Assumption~1 is reasonable. With Assumption~1, under any constant reference input $v(t) \equiv r$, the associated steady-state values of $\bar{x}(t)$ and $y(t)$, denoted as
\begin{subequations}
\begin{align}
\bar{x}^*(r) &= \begin{bmatrix}
x_p^*(r) \\ x_u^*(r)
\end{bmatrix} = (I-\bar{A})^{-1} \bar{B}_v r, \\
y^*(r) &= \left(\bar{C}(I-\bar{A})^{-1} \bar{B}_v + \bar{D}_v \right) r,
\end{align}
\end{subequations}
are exponentially stable in the disturbance-free case (i.e., with $w(t) \equiv 0$).

{\it Assumption 2:} The disturbance inputs $\{w(t)\}_{t \in \mathbb{Z}_{\ge 0}}$ are independent and identically distributed ({\it i.i.d}) Gaussian random variables with zero mean and covariance matrix $W$. They are also independent of $\bar{x}(0)$ and $\{v(t)\}_{t \in \mathbb{Z}_{\ge 0}}$. We denote such disturbance inputs as
\begin{equation}
\{w(t)\}_{t \in \mathbb{Z}_{\ge 0}} \sim \mathcal{N}(0,W).
\end{equation}

The system is assumed to be subject to the following set of linear inequality constraints for all $t \in \mathbb{Z}_{\ge 0}$,
\begin{equation}\label{equ:deterministic_constraint}
G_i^{\top} y(t) \le g_i, \quad i = 1,...,n_g,
\end{equation}
where $G_i \in \mathbb{R}^{n_y}$ and $g_i \in \mathbb{R}$. Because the Gaussian disturbance inputs $\{w(t)\}_{t \in \mathbb{Z}_{\ge 0}} \sim \mathcal{N}(0,W)$ are not compactly supported, it is generally not possible to enforce the constraints \eqref{equ:deterministic_constraint} deterministically. Instead, we are interested in enforcing them with high probabilities. In particular, we consider the following set of chance constraints,
\begin{equation}\label{equ:chance_constraint}
\mathbb{P} \left\{G_i^{\top} y(t) \le g_i \right\} \ge \beta_i, \quad i = 1,...,n_g,
\end{equation}
with $\beta_i \in (0.5,1)$ being specified confidence levels of constraint satisfaction.

In what follows, we introduce a controller state and reference governor (CSRG) scheme to enforce \eqref{equ:chance_constraint}. 

\section{Controller State and Reference Governor}\label{sec:3}

The CSRG is an add-on scheme for the closed-loop system \eqref{equ:system_cpt} that supervises the controller state $x_u(t)$ and the reference input $v(t)$ to enforce the constraints \eqref{equ:chance_constraint}. As an extension of both the deterministic CSRG in \cite{mcdonough2015controller} and the stochastic RG in \cite{kalabic2019reference}, our stochastic CSRG utilizes the following maximal output admissible set,
\begin{equation}\label{equ:O_all}
\mathcal{O}_{\infty} = \bigcap_{i = 1}^{n_g} \mathcal{O}_{\infty}^i,
\end{equation}
where $\mathcal{O}_{\infty}^i$ is defined as
\begin{align}\label{equ:O_i}
& \mathcal{O}_{\infty}^i = \big\{(x_p,x_u,v) \in \mathbb{R}^{n_{xp}} \!\times\! \mathbb{R}^{n_{xu}} \!\times\! \mathbb{R}^{n_v}:\, \text{if } x_p(0) = x_p, \nonumber \\
&\quad x_u(0) = x_u, v(t) \equiv v, \text{ and } \{w(t)\}_{t \in \mathbb{Z}_{\ge 0}} \sim \mathcal{N}(0,W), \nonumber \\
&\quad \text{then } \mathbb{P} \big\{G_i^{\top} y(t) \le g_i \big\} \ge \beta_i \text{ for all } t \in \mathbb{Z}_{\ge 0} \big\}.
\end{align}

With $\mathcal{O}_{\infty}$, the CSRG determines the values of $x_u(t)$ and $v(t)$ based on the following constrained optimization problem,
\begin{subequations}\label{equ:opt_1}
\begin{align}
\min_{x_u,v} &\quad J\left(x_u,v,x_p(t),r(t)\right), \\
\text{subject to} &\quad \left(x_p(t),x_u,v\right) \in \mathcal{O}_{\infty},
\end{align}
\end{subequations}
with the cost function defined as
\begin{align}\label{equ:cost}
& J\left(x_u,v,x_p(t),r(t)\right) \nonumber \\
&= \left\|\begin{bmatrix}
x_p(t) \\ x_u
\end{bmatrix} - \begin{bmatrix}
x_p^*(v) \\ x_u^*(v)
\end{bmatrix}\right\|_{P}^2 + \left\|v - r(t)\right\|_{R}^2 \nonumber \\
&= \left\|\begin{bmatrix}
x_p(t) \\ x_u
\end{bmatrix} - (I-\bar{A})^{-1} \bar{B}_v v \right\|_{P}^2 + \left\|v - r(t)\right\|_{R}^2,
\end{align}
where $x_p(t)$ is the current plant state value, $r(t)$ denotes the commanded/nominal value of the reference input, typically provided by a human operator or generated by a higher-level planning algorithm, and $\|\cdot\|_P = \sqrt{(\cdot)^{\top}P(\cdot)}$, $\|\cdot\|_R = \sqrt{(\cdot)^{\top}R(\cdot)}$ with $P,R$ being positive-definite matrices. In particular, the matrix $P$ is selected as the positive-definite solution to the Lyapunov equation,
\begin{equation}\label{equ:P_matrix}
\bar{A}^{\top} P \bar{A} - P + Q = 0,
\end{equation}
with $Q$ being a positive-definite matrix. Note that the cost function \eqref{equ:cost} is convex quadratic in the decision variables $(x_u,v)$.

\subsection{Maximal output admissible set}\label{sec:31}

The CSRG enforces the chance constraints \eqref{equ:chance_constraint} using the maximal output admissible set $\mathcal{O}_{\infty}$ defined by \eqref{equ:O_all} and \eqref{equ:O_i}. In this section, we address the
explicit expression and algorithmic determination of $\mathcal{O}_{\infty}$.

Given $\bar{x}(0) = [x_p^{\top}(0), x_u^{\top}(0)]^{\top}$, $v(t) \equiv v$ and $\{w(t)\}_{t \in \mathbb{Z}_{\ge 0}} \sim \mathcal{N}(0,W)$, the outputs of \eqref{equ:system_cpt} are normally distributed according to
\begin{equation}\label{equ:yt}
y(t) \sim \mathcal{N}\left(\bar{y}(t),\Sigma_y(t)\right),
\end{equation}
where the mean $\bar{y}(t)$ is determined as
\begin{equation}\label{equ:mean}
\bar{y}(t) = \bar{C} \bar{A}^t \bar{x}(0) + \left(\bar{C} \sum_{k=0}^{t-1} \bar{A}^k \bar{B}_v + \bar{D}_v\right) v,
\end{equation}
and the covariance $\Sigma_y(t)$ satisfies
\begin{subequations}\label{equ:covariance}
\begin{align}
\Sigma_x(t+1) &= \bar{A} \Sigma_x(t) \bar{A}^{\top} + \bar{B}_w W \bar{B}_w^{\top}, \\
\Sigma_y(t) &= \bar{C} \Sigma_x(t) \bar{C}^{\top} + \bar{D}_w W \bar{D}_w^{\top},
\end{align}
\end{subequations}
with $\Sigma_x(0) = 0$. From \eqref{equ:covariance} and $\Sigma_x(0) = 0$, one can also derive the following explicit expression for $\Sigma_y(t)$,
\begin{equation}\label{equ:covariance_2}
\Sigma_y(t) =  \bar{C} \left(\sum_{k = 0}^{t-1} \bar{A}^k \bar{B}_w W \bar{B}_w^{\top} (\bar{A}^{\top})^k \right) \bar{C}^{\top} + \bar{D}_w W \bar{D}_w^{\top}.
\end{equation}
Based on \eqref{equ:yt}, $G_i^{\top} y(t) \in \mathbb{R}$ is normally distributed according to
\begin{equation}\label{equ:Gyt}
G_i^{\top} y(t) \sim \mathcal{N}\left(G_i^{\top} \bar{y}(t), G_i^{\top} \Sigma_y(t) G_i\right).
\end{equation}
Then, the probability $\mathbb{P} \left\{G_i^{\top} y(t) \le g_i \right\}$ can be computed as
\begin{align}
\mathbb{P} \left\{G_i^{\top} y(t) \le g_i\right\} &= \Phi \left(\frac{g_i - G_i^{\top}\bar{y}(t)}{\sqrt{G_i^{\top} \Sigma_y(t) G_i}} \right) \\
&= \frac{1}{2}\left[1+\text{erf}\left(\frac{g_i - G_i^{\top}\bar{y}(t)}{\sqrt{2G_i^{\top} \Sigma_y(t) G_i}}\right)\right], \nonumber
\end{align}
where $\Phi(\cdot)$ denotes the cumulative distribution function of the standard normal distribution and $\text{erf}(\cdot)$ denotes the error function
\begin{equation}
\text{erf}(z) = \frac{2}{\sqrt{\pi}} \int_{0}^z e^{-t^2}\, \text{d}t.
\end{equation}
Note that $\Phi(\cdot)$ and $\text{erf}(\cdot)$ are related according to $\Phi(z) = \frac{1}{2} \left[1+\text{erf}\left(\frac{z}{\sqrt{2}}\right)\right]$.

Then, the chance constraint \eqref{equ:chance_constraint} becomes
\begin{equation}
\frac{1}{2}\left[1+\text{erf}\left(\frac{g_i - G_i^{\top}\bar{y}(t)}{\sqrt{2G_i^{\top} \Sigma_y(t) G_i}}\right)\right] \ge \beta_i,
\end{equation}
which can be equivalently written as
\begin{align}\label{equ:chance_constraint_erf}
G_i^{\top} \bar{y}(t) &= G_i^{\top} \bar{C} \bar{A}^t \begin{bmatrix} x_p(0) \\ x_u(0) \end{bmatrix} + G_i^{\top} \left(\bar{C} \sum_{k=0}^{t-1} \bar{A}^k \bar{B}_v + \bar{D}_v\right) v \nonumber \\
&\le g_i - \sqrt{2G_i^{\top} \Sigma_y(t) G_i}\, \text{erf}^{-1} \big(2\beta_i - 1\big),
\end{align}
where $\text{erf}^{-1}(\cdot)$ denotes the inverse of the error function $\text{erf}(\cdot)$. Note that the $\Sigma_y(t)$ in \eqref{equ:chance_constraint_erf}, computed from \eqref{equ:covariance}, is independent of $(x_p(0),x_u(0),v)$, and accordingly, \eqref{equ:chance_constraint_erf} is a linear inequality condition on the triple $(x_p(0),x_u(0),v)$.

With \eqref{equ:chance_constraint_erf}, the set $\mathcal{O}_{\infty}^i$ defined in \eqref{equ:O_i} can be explicitly expressed as
\begin{align}\label{equ:O_i2}
& \mathcal{O}_{\infty}^i = \Bigg\{(x_p,x_u,v): G_i^{\top} \bar{C} \bar{A}^t \begin{bmatrix} x_p \\ x_u \end{bmatrix} + G_i^{\top} \bigg(\bar{C} \sum_{k=0}^{t-1} \bar{A}^k \bar{B}_v + \nonumber \\
&\bar{D}_v\bigg) v \le g_i - \sqrt{2G_i^{\top} \Sigma_y(t) G_i}\, \text{erf}^{-1} \big(2\beta_i - 1\big), \forall t \in \mathbb{Z}_{\ge 0}\Bigg\}.
\end{align}
The set $\mathcal{O}_{\infty}^i$ in \eqref{equ:O_i2} is characterized by an infinite number of linear inequalities ($\forall t \in \mathbb{Z}_{\ge 0}$) and cannot be finitely determined in general, and so is $\mathcal{O}_{\infty} = \bigcap_{i = 1}^{n_g} \mathcal{O}_{\infty}^i$. Hence, we consider inner approximations of $\mathcal{O}_{\infty}^i$ and $\mathcal{O}_{\infty}$, denoted by $\tilde{\mathcal{O}}_{\infty}^i$ and $\tilde{\mathcal{O}}_{\infty}$ respectively, which are defined as
\begin{subequations}\label{equ:O_all2}
\begin{align}
\tilde{\mathcal{O}}_{\infty}^i &= \mathcal{O}_{\infty}^i \cap \left(\mathbb{R}^{n_{xp}} \times \mathbb{R}^{n_{xu}} \times \tilde{\Omega}^i \right), \\
\tilde{\mathcal{O}}_{\infty} &= \bigcap_{i = 1}^{n_g} \tilde{\mathcal{O}}_{\infty}^i,
\end{align}
\end{subequations}
where $\tilde{\Omega}^i$ is a compact and convex subset of $\text{int}(\Omega^i)$ with nonempty interior. The set $\Omega^i \subset \mathbb{R}^{n_v}$ is defined as
\begin{align}\label{equ:omega_i}
\Omega^i =&\, \bigg\{ v \in \mathbb{R}^{n_v}:\, G_i^{\top} \left(\bar{C}(I-\bar{A})^{-1} \bar{B}_v + \bar{D}_v \right) v \nonumber \\
&\, \le g_i - \sqrt{2G_i^{\top} \Sigma_y^{\infty} G_i}\, \text{erf}^{-1} \big(2\beta_i - 1\big)\bigg\},
\end{align}
if $G_i^{\top} \left(\bar{C}(I-\bar{A})^{-1} \bar{B}_v + \bar{D}_v \right) \neq 0$, and $\Omega^i = \mathbb{R}^{n_v}$ otherwise. In \eqref{equ:omega_i}, $\Sigma_y^{\infty}$ denotes the steady-state covariance of $y$, computed from
\begin{equation}\label{equ:steady_cov_y}
\Sigma_y^{\infty} = \bar{C} \Sigma_x^{\infty} \bar{C}^{\top} + \bar{D}_w W \bar{D}_w^{\top},
\end{equation}
where $\Sigma_x^{\infty}$ denotes the steady-state covariance of $\bar{x}$, i.e., $\Sigma_x^{\infty} = \lim_{t \to \infty}\Sigma_x(t)$, and is obtained as the solution to the following Lyapunov equation,
\begin{equation}\label{equ:Lyapunov_2}
\Sigma_x^{\infty} = \bar{A} \Sigma_x^{\infty} \bar{A}^{\top} + \bar{B}_w W \bar{B}_w^{\top}.
\end{equation}
Note that $\lim_{t \to \infty}\Sigma_x(t)$ exists and is equal to the solution of \eqref{equ:Lyapunov_2}, which is guaranteed by $\bar{A}$ being Schur (see Assumption~1). For the latter case with $G_i^{\top} \left(\bar{C}(I-\bar{A})^{-1} \bar{B}_v + \bar{D}_v \right) = 0$, we assume $g_i - \sqrt{2G_i^{\top} \Sigma_y^{\infty} G_i}\, \text{erf}^{-1} \big(2\beta_i - 1\big) > 0$.

Algorithmically, the inner approximation $\tilde{\mathcal{O}}_{\infty}$ of $\mathcal{O}_{\infty}$ is constructed through the following recursions,
\begin{equation}\label{equ:O_t_1}
\tilde{\mathcal{O}}_{t+1} = \tilde{\mathcal{O}}_{t} \cap \Xi_{t+1},
\end{equation}
where $\tilde{\mathcal{O}}_{0} = \Xi_{0} \cap \left(\mathbb{R}^{n_{xp}} \times \mathbb{R}^{n_{xu}} \times \bigcap_{i=1}^{n_g} \tilde{\Omega}^i\right)$, and
\begin{align}\label{equ:O_t_2}
\Xi_t &= \Bigg\{(x_p,x_u,v): G^{\top}\bar{C} \bar{A}^t \begin{bmatrix} x_p \\ x_u \end{bmatrix} + G^{\top}\bigg(\bar{C} \sum_{k=0}^{t-1} \bar{A}^k \bar{B}_v + \nonumber \\
& \bar{D}_v\bigg) v \le g - \begin{bmatrix} \sqrt{2G_1^{\top} \Sigma_y(t) G_1}\, \text{erf}^{-1} \big(2\beta_1 - 1\big) \\ \vdots \\  \sqrt{2G_{n_g}^{\top} \Sigma_y(t) G_{n_g}}\, \text{erf}^{-1} \big(2\beta_{n_g} - 1\big) \end{bmatrix} \Bigg\},
\end{align}
where $G \in \mathbb{R}^{n_y \times n_g}$ is the matrix with $G_i$, $i = 1,...,n_g$, as its $i$th column, and $g \in \mathbb{R}^{n_g}$ is the vector with $g_i$ as its $i$th entry.

The set $\tilde{\mathcal{O}}_{\infty}$ is finitely determined. This is formally presented as the following proposition:

{\bf Proposition 1:} Suppose $(\bar{C},\bar{A})$ is observable and the set $\mathcal{Y} \subset \mathbb{R}^{n_y}$ defined by the constraints \eqref{equ:deterministic_constraint} is bounded. Then, (i) there exists $t^* \in \mathbb{Z}_{\ge 0}$ such that $\tilde{\mathcal{O}}_{\infty} = \tilde{\mathcal{O}}_{t^*}$, and (ii) $\tilde{\mathcal{O}}_{\infty}$ is compact and convex.

{\bf Proof:} For any $v \in \tilde{\Omega}^i \subset \text{int}(\Omega^i)$, there exists $\varepsilon_i(v)>0$ such that $\mathcal{B}_{2\varepsilon_i(v)}(v) \subset \Omega^i$, where $\mathcal{B}_{2\varepsilon_i(v)}(v)$ denotes the open ball in $\mathbb{R}^{n_v}$ centered at $v$ with radius $2\varepsilon_i(v)$. Since $\tilde{\Omega}^i$ is compact and $\big\{\mathcal{B}_{\varepsilon_i(v)}(v): v \in \tilde{\Omega}^i\big\}$ is an open cover of $\tilde{\Omega}^i$, there is a finite subcover $\big\{\mathcal{B}_{\varepsilon_i(v_j)}(v_j): j = 1,...,J_i\big\} \subset \big\{\mathcal{B}_{\varepsilon_i(v)}(v): v \in \tilde{\Omega}^i\big\}$. This means for any $v \in \tilde{\Omega}^i$, we have $v \in \mathcal{B}_{\varepsilon_i(v_j)}(v_j)$ for some $j = 1,...,J_i$. Let $\varepsilon_i = \min \big\{\varepsilon_i(v_j): j = 1,...,J_i \big\}$. Then, for any $v \in \tilde{\Omega}^i$, we have $v \oplus \mathcal{B}_{\varepsilon_i}(0) \subset \mathcal{B}_{\varepsilon_i(v_j)}(v_j) \oplus \mathcal{B}_{\varepsilon_i(v_j)}(0) = \mathcal{B}_{2\varepsilon_i(v_j)}(v_j) \subset \Omega^i$. In particular, $v$ satisfies
\begin{align}\label{equ:P1_1}
& G_i^{\top} \left(\bar{C}(I-\bar{A})^{-1} \bar{B}_v + \bar{D}_v \right) v + \bar{\varepsilon}_i \nonumber \\
&\, \le g_i - \sqrt{2G_i^{\top} \Sigma_y^{\infty} G_i}\, \text{erf}^{-1} \big(2\beta_i - 1\big),
\end{align}
where $\bar{\varepsilon}_i = \sup_{v \in \mathcal{B}_{\varepsilon_i}(0)} G_i^{\top} \left(\bar{C}(I-\bar{A})^{-1} \bar{B}_v + \bar{D}_v \right) v > 0$ if $G_i^{\top} \big(\bar{C}(I-\bar{A})^{-1} \bar{B}_v + \bar{D}_v \big) \neq 0$ and $\bar{\varepsilon}_i = g_i - \sqrt{2G_i^{\top} \Sigma_y^{\infty} G_i}\, \text{erf}^{-1} \big(2\beta_i - 1\big) > 0$ otherwise. 

Because $(\bar{C},\bar{A})$ is observable and $\mathcal{Y}$ and $\bigcap_{i=1}^{n_g} \tilde{\Omega}^i$ are bounded, there exists $t_0 \in \mathbb{Z}_{\ge 0}$ such that $\tilde{\mathcal{O}}_{t_0}$ constructed according to \eqref{equ:O_t_1} is bounded. Then, because $\bar{A}$ is Schur, there exists $t_1 \in \mathbb{Z}_{\ge t_0}$ such that for any $(x_p,x_u,v) \in \tilde{\mathcal{O}}_{t_0}$, we have
\begin{subequations}\label{equ:P1_2}
\begin{align}
& G_i^{\top}\bar{C} \bar{A}^t \begin{bmatrix} x_p \\ x_u \end{bmatrix} \le \frac{\bar{\varepsilon}_i}{3}, \\
& G_i^{\top}\! \left(\!\bar{C} \sum_{k=0}^{t-1} \bar{A}^k \bar{B}_v \!+\! \bar{D}_v\!\right) v - G_i^{\top} \left(\bar{C}(I-\bar{A})^{-1} \bar{B}_v \!+\! \bar{D}_v \right) v \nonumber \\
&= -G_i^{\top} \bar{C} \bar{A}^t (I-\bar{A})^{-1} \bar{B}_v v \le \frac{\bar{\varepsilon}_i}{3}, \\
& \left(\sqrt{2G_i^{\top} \Sigma_y(t) G_i} - \sqrt{2G_i^{\top} \Sigma_y^{\infty} G_i}\, \right) \text{erf}^{-1} \big(2\beta_i \!-\! 1\big) \le \frac{\bar{\varepsilon}_i}{3},
\end{align}
\end{subequations}
for $i = 1,...,n_g$ and all $t \in \mathbb{Z}_{\ge t_1}$. Combining \eqref{equ:P1_1} and \eqref{equ:P1_2}, we obtain
\begin{align}
& G_i^{\top}\bar{C} \bar{A}^t \begin{bmatrix} x_p \\ x_u \end{bmatrix} + G_i^{\top} \left(\bar{C} \sum_{k=0}^{t-1} \bar{A}^k \bar{B}_v + \bar{D}_v\right) v \nonumber \\
&\le G_i^{\top} \left(\bar{C}(I-\bar{A})^{-1} \bar{B}_v + \bar{D}_v \right) v + \frac{2\bar{\varepsilon}_i}{3} \nonumber \\
&\le g_i - \sqrt{2G_i^{\top} \Sigma_y^{\infty} G_i}\, \text{erf}^{-1} \big(2\beta_i - 1\big) - \frac{\bar{\varepsilon}_i}{3} \nonumber \\
&\le g_i - \sqrt{2G_i^{\top} \Sigma_y(t) G_i}\, \text{erf}^{-1} \big(2\beta_i - 1\big),
\end{align}
for $i = 1,...,n_g$ and all $t \in \mathbb{Z}_{\ge t_1}$. This means for any $(x_p,x_u,v) \in \tilde{\mathcal{O}}_{t_0}$, we have $(x_p,x_u,v) \in \Xi_t$ for all $t \in \mathbb{Z}_{\ge t_1}$, i.e., $\tilde{\mathcal{O}}_{t_0} \subset \Xi_t$ for all $t \in \mathbb{Z}_{\ge t_1}$. Therefore, we have
\begin{align}
\tilde{\mathcal{O}}_{\infty} &= \tilde{\mathcal{O}}_{t_0} \cap \bigcap_{t = t_0}^{t_1-1} \Xi_t \cap \bigcap_{t = t_1}^{\infty} \Xi_t \supset \tilde{\mathcal{O}}_{t_0} \cap \bigcap_{t = t_0}^{t_1-1} \Xi_t \cap \tilde{\mathcal{O}}_{t_0} \nonumber \\
&= \tilde{\mathcal{O}}_{t_0} \cap \bigcap_{t = t_0}^{t_1-1} \Xi_t = \tilde{\mathcal{O}}_{t_1-1}.
\end{align}
Meanwhile, we also have $\tilde{\mathcal{O}}_{\infty} = \tilde{\mathcal{O}}_{t_1-1} \cap \bigcap_{t = t_1}^{\infty} \Xi_t \subset \tilde{\mathcal{O}}_{t_1-1}$, which yields $\tilde{\mathcal{O}}_{\infty} = \tilde{\mathcal{O}}_{t^*}$ with $t^* = t_1-1 \in \mathbb{Z}_{\ge 0}$. This proves (i).

For (ii), because $\tilde{\mathcal{O}}_{\infty} = \tilde{\mathcal{O}}_{t_0} \cap \bigcap_{t = t_0}^{\infty} \Xi_t$ and $\tilde{\mathcal{O}}_{t_0}$ is bounded, $\tilde{\mathcal{O}}_{\infty}$ is also bounded. Because $\tilde{\mathcal{O}}_{\infty} = \big(\bigcap_{t = 0}^{\infty} \Xi_t\big) \cap \big(\mathbb{R}^{n_{xp}} \times \mathbb{R}^{n_{xu}} \times \bigcap_{i=1}^{n_g} \tilde{\Omega}^i\big)$ where $\Xi_t$, $t \in \mathbb{Z}_{\ge 0}$, and $\mathbb{R}^{n_{xp}} \times \mathbb{R}^{n_{xu}} \times \bigcap_{i=1}^{n_g} \tilde{\Omega}^i$ are closed and convex, $\tilde{\mathcal{O}}_{\infty}$ is also closed and convex. This proves the compactness and convexity of $\tilde{\mathcal{O}}_{\infty}$. $\blacksquare$

\subsection{CSRG algorithms}\label{sec:32}

In this section, we present two CSRG algorithms. The algorithms are based on the constrained optimization problem \eqref{equ:opt_1} and augmented with additional features to achieve improved performance.

{\it Algorithm 1:} At each discrete-time instant $t \in \mathbb{Z}_{\ge 0}$,
\begin{align}\label{equ:opt_2}
& \left(x_u(t),v(t)\right) = \\
& \begin{cases} \left(\hat{x}_u(t),\hat{v}(t)\right), &\!\! \text{ if } \left(x_p(t),\bar{x}_u(t),v(t-1)\right) \in \tilde{\mathcal{O}}_{\infty} \\
&\!\! \text{ and } \eqref{equ:opt_3} \text{ is feasible}, \\[2pt]
\left(\bar{x}_u(t),v(t-1)\right), &\!\! \text{ otherwise}, \end{cases} \nonumber
\end{align}
where $\bar{x}_u(t) = A_p x_p(t-1) + A_u x_u(t-1) + D_v v(t-1)$, and the pair $\left(\hat{x}_u(t),\hat{v}(t)\right)$ is determined as
\begin{subequations}\label{equ:opt_3}
\begin{align}
\left(\hat{x}_u(t),\hat{v}(t)\right) = \argmin_{x_u,v} &\quad J\left(x_u,v,x_p(t),r(t)\right), \label{equ:opt_31} \\
\text{subject to} &\quad \left(x_p(t),x_u,v\right) \in \tilde{\mathcal{O}}_{\infty}, \label{equ:opt_32} \\[7pt]
&\!\!\!\!\!\!\!\!\!\!\!\!\!\!\!\!\!\!\!\!\!\!\!\!\!\!\!\!\!\!\!\!\!\!\!\!\!\!\!\!\!\!\!\! \|v - r(t)\|_{R}^2 \le \left(\max \left\{\|v(t-1) - r(t)\|_{R} - \delta, 0\right\}\right)^2, \label{equ:opt_33}
\end{align}
\end{subequations}
with $\delta>0$ being a sufficiently small constant.

We remark that unlike in the deterministic case \citep{mcdonough2015controller,mcdonough2015integrator}, recursive feasibility of the constrained optimization problem \eqref{equ:opt_3} cannot be guaranteed in our considered stochastic setting. Specifically, because the Gaussian disturbance variables $\{w(t)\}_{t \in \mathbb{Z}_{\ge 0}} \sim \mathcal{N}(0,W)$ are not compactly supported, there is always a non-zero probability for the plant state $x_p(t)$ to get outside of ${\rm proj}_{x_p}(\tilde{\mathcal{O}}_\infty)$, which denotes the projection of the compact set $\tilde{\mathcal{O}}_\infty \in \mathbb{R}^{n_{xp}} \times \mathbb{R}^{n_{xu}} \times \mathbb{R}^{n_v}$ onto the $x_p$-space. Therefore, we introduce an infeasibility-handling mechanism in \eqref{equ:opt_2}. In particular, when $\left(x_p(t),\bar{x}_u(t),v(t-1)\right) \notin \tilde{\mathcal{O}}_{\infty}$ or \eqref{equ:opt_3} is infeasible, the controller state and reference pair $\left(x_u(t),v(t)\right)$ is set to $\left(\bar{x}_u(t),v(t-1)\right)$, which was previously determined to be chance-constraint admissible. Moreover, it has been observed in \cite{kalabic2019reference} that if the direction of reference modification is not restricted, a chance-constrained RG may move $v(t)$ away from $r(t)$, even cause sudden large changes in $v(t)$, to enforce constraints. This can impede the convergence of $v(t)$ to $r(t)$ and degrade system performance. Therefore, $\|v - r(t)\|_{R}^2 \le \left(\max \left\{\|v(t-1) - r(t)\|_{R} - \delta, 0\right\}\right)^2$ is added to the optimization problem \eqref{equ:opt_3} as an extra constraint, \eqref{equ:opt_33}, to enforce $\hat{v}(t)$ to be closer to $r(t)$ than the previous reference value $v(t-1)$. In particular, \eqref{equ:opt_33} is a convex quadratic constraint on the decision variable $v$.




In some circumstances, the convergence of the modified reference $v(t)$ to the commanded value $r(t)$ is prioritized so that the system can potentially have faster response to human operator intention or planning algorithm schedule. For this, the following Algorithm~2 can be used.

{\it Algorithm 2:} At each discrete-time instant $t \in \mathbb{Z}_{\ge 0}$,
\begin{align}\label{equ:opt_4}
& \left(x_u(t),v(t)\right) =  \\
& \begin{cases}\,\, \left(\hat{x}_u'(t),r(t)\right),\,\,\,\, \text{ if } v(t-1) \neq r(t) \text{ and \eqref{equ:opt_5} is feasible}, \\
\text{ the solution to } \eqref{equ:opt_2},\,\,\,\, \text{ otherwise}, \end{cases} \nonumber 
\end{align}
where $\hat{x}_u'(t)$ is determined as
\begin{subequations}\label{equ:opt_5}
\begin{align}
\hat{x}_u'(t) = \argmin_{x_u} &\,\,\,\, J\left(x_u,r(t),x_p(t),r(t)\right), \\
\text{subject to} &\,\,\,\, \left(x_p(t),x_u,r(t)\right) \in \tilde{\mathcal{O}}_{\infty}. \label{equ:opt_51}
\end{align}
\end{subequations}

In what follows, we characterize theoretical properties of our CSRG algorithms, and also extend the CSRG scheme to treat joint chance constraints.

\section{Theoretical Properties and Extension to Joint Chance Constraints}\label{sec:4}

In this section, we first derive theoretical properties of our CSRG algorithms. Moreover, the CSRG scheme introduced in Section~\ref{sec:3} uses the maximal output admissible set $\tilde{\mathcal{O}}_{\infty}$ to enforce the set of {\it individual chance constraints} \eqref{equ:chance_constraint}. In the second part of this section, we extend this nominal scheme to treat {\it joint chance constraints}.

\subsection{Theoretical properties}

Three important properties of the CSRG scheme are: 1) constraint enforcement, 2) convergence of the modified reference $v(t)$ to constant, steady-state constraint-admissible commanded value $r$, and 3) stability of the commanded set-point $\bar{x}^*(r)$. The first two are also important properties of conventional RGs \citep{garone2017reference,kalabic2019reference}. Unlike the conventional RG scheme where stability of $\bar{x}^*(r)$ is inherited from stability of the nominal closed-loop system, for the CSRG scheme stability needs to be separately verified, because the CSRG scheme also modifies the control input signal $u(t)$ through modifying the controller state $x_u(t)$. 

Due to the presence of stochastic disturbances, the techniques for establishing the above three properties of the chance-constrained CSRG scheme are significantly different from the techniques used in \cite{mcdonough2015controller} for the deterministic case. For instance, in the deterministic case these properties are established based on the positive invariance of the maximal output admissible set $\tilde{\mathcal{O}}_{\infty}$. In contrast, in our considered stochastic setting the set $\tilde{\mathcal{O}}_{\infty}$ is not positively invariant, as has been discussed above, and consequently, new techniques for establishing these properties need to be developed.


Firstly, the following proposition establishes the chance-constraint enforcement property of our CSRG algorithms. The significance is that with the infeasibility-handling mechanism that we have introduced, the desired probabilistic constraint enforcement guarantee is maintained in closed-loop operation of the system.

{\bf Proposition 2:} Suppose $\left(x_p(0),x_u(0),v(0)\right) \in \tilde{\mathcal{O}}_{\infty}$. Then, the closed-loop response of the overall system (consisting of the plant \eqref{equ:system_plant}, the nominal controller \eqref{equ:system_controller}, and the CSRG algorithm~1 or 2) satisfies the chance constraints \eqref{equ:chance_constraint} for all $t \in \mathbb{Z}_{\ge 0}$. 

{\bf Proof:} Let $t \in \mathbb{Z}_{\ge 0}$ be arbitrary and define $\tau = \max \big\{t' \in \mathbb{Z}_{[0,t]}: \left(x_p(t'),x_u(t'),v(t')\right) \in \tilde{\mathcal{O}}_{\infty}\big\}$. Since it is assumed that $\left(x_p(0),x_u(0),v(0)\right) \in \tilde{\mathcal{O}}_{\infty}$, the set $\big\{t' \in \mathbb{Z}_{[0,t]}: \left(x_p(t'),x_u(t'),v(t')\right) \in \tilde{\mathcal{O}}_{\infty}\big\}$ is necessarily non-empty and the variable $\tau$ is therefore well-defined. Note that $\tau$ is a random variable with the finite support $\mathbb{Z}_{[0,t]}$. In particular, we have
\begin{equation}\label{equ:P21}
  \sum_{k = 0}^{t} \mathbb{P} \left\{\tau = k \right\} = \mathbb{P} \left\{\tau \in \mathbb{Z}_{[0,t]} \right\} = 1.
\end{equation}
If $\tau = k$, then according to our CSRG algorithms, the trajectory of controller state and reference pair must satisfy $\left(x_u(t'),v(t')\right) = \left(\bar{x}_u(t'),v(k)\right)$ for all $t' \in \mathbb{Z}_{[k+1,t]}$. In this case, the plant and controller states $\bar{x}(t') = [x_p^{\top}(t'), x_u^{\top}(t')]^{\top}$ evolve according to \eqref{equ:system_cpt_1} with $v(t') \equiv v(k)$ over $t' \in \mathbb{Z}_{[k,t]}$. Since $\left(x_p(k),x_u(k),v(k)\right) \in \tilde{\mathcal{O}}_{\infty} \subset \mathcal{O}_{\infty}^i$ and the dynamics of $\bar{x}(t') = [x_p^{\top}(t'), x_u^{\top}(t')]^{\top}$ follow \eqref{equ:system_cpt_1} with $v(t') \equiv v(k)$ for $t' \in \mathbb{Z}_{[k,t]}$, using the definition of $\mathcal{O}_{\infty}^i$ in \eqref{equ:O_i}, we have
\begin{equation}\label{equ:P22}
   \mathbb{P} \left\{G_i^{\top} y(t) \le g_i \big|\tau = k \right\} \ge \beta_i.
\end{equation}
Note that this holds for all $k = 0,...,t$ and all $i = 1,...,n_g$.

Combining \eqref{equ:P21} and \eqref{equ:P22}, we obtain
\begin{align}
  \mathbb{P} \left\{G_i^{\top} y(t) \le g_i \right\} &= \sum_{k = 0}^{t} \mathbb{P} \left\{G_i^{\top} y(t) \le g_i \big| \tau = k \right\} \mathbb{P} \left\{\tau = k \right\} \nonumber \\
  &\ge \beta_i \sum_{k = 0}^{t} \mathbb{P} \left\{\tau = k \right\} = \beta_i,
\end{align}
for $i = 1,...,n_g$. As $t \in \mathbb{Z}_{\ge 0}$ is arbitrary, the result follows. $\blacksquare$

We next discuss the convergence property of the modified reference $v(t)$ to commanded value $r(t)$. To begin with, we introduce the following lemma, which will be used to prove the main convergence result, Proposition~3.

{\bf Lemma 1:} Given $\tilde{\mathcal{O}}_{\infty}$, there exists $\varepsilon>0$ such that $\mathcal{B}_{\varepsilon}\left(\bar{x}^*(v)\right) \times \{v\} \subset \tilde{\mathcal{O}}_{\infty}$ for any $v \in \bigcap_{i=1}^{n_g} \tilde{\Omega}^i$, where $\mathcal{B}_{\varepsilon}\left(\bar{x}^*(v)\right)$ denotes the open ball in $\mathbb{R}^{n_{xp}+n_{xu}}$ centered at $\bar{x}^*(v)$ with radius $\varepsilon$.

{\bf Proof:} Firstly, it is easily seen from \eqref{equ:covariance_2} that for any $t \in \mathbb{Z}_{\ge 0}$, $\Sigma_y(t+1) - \Sigma_y(t) = (\bar{C} \bar{A}^t \bar{B}_w) W (\bar{C} \bar{A}^t \bar{B}_w)^{\top}$ is positive semi-definite, denoted as $\Sigma_y(t+1) \succeq \Sigma_y(t)$. As a result, it holds that $\Sigma_y^{\infty} \succeq \Sigma_y(t)$ for all $t \in \mathbb{Z}_{\ge 0}$, where $\Sigma_y^{\infty}$ denotes the steady-state covariance of $y$, computed from \eqref{equ:steady_cov_y}. In turn, $\sqrt{2G_i^{\top} \Sigma_y^{\infty} G_i} \ge \sqrt{2G_i^{\top} \Sigma_y(t) G_i}$ for all $t \in \mathbb{Z}_{\ge 0}$ and $i = 1,...,n_g$.

Referring to \eqref{equ:P1_1} in the proof of Proposition~1, for each $i = 1,...,n_g$, there exists $\bar{\varepsilon}_i > 0$ such that for any $v \in \tilde{\Omega}^i$,
\begin{align}\label{equ:LM1_1}
& G_i^{\top}\bar{C} \bar{A}^t \bar{x}^*(v) + G_i^{\top} \left(\bar{C} \sum_{k=0}^{t-1} \bar{A}^k \bar{B}_v + \bar{D}_v\right) v \nonumber \\
&= G_i^{\top} \left(\bar{C}(I-\bar{A})^{-1} \bar{B}_v + \bar{D}_v \right) v + \bar{\varepsilon}_i \nonumber \\
&\le g_i - \sqrt{2G_i^{\top} \Sigma_y^{\infty} G_i}\, \text{erf}^{-1} \big(2\beta_i - 1\big) \nonumber \\
&\le g_i - \sqrt{2G_i^{\top} \Sigma_y(t) G_i}\, \text{erf}^{-1} \big(2\beta_i - 1\big),
\end{align}
for all $t \in \mathbb{Z}_{\ge 0}$.

Now let $\varepsilon = \min \left\{\frac{\bar{\varepsilon}_i}{\sup_{t \ge 0} \|G_i^{\top}\bar{C} \bar{A}^t\|}: i = 1,...,n_g \right\}$ and obtain that for any $v \in \bigcap_{i=1}^{n_g} \tilde{\Omega}^i$ and $\bar{x} \in \mathcal{B}_{\varepsilon}\left(\bar{x}^*(v)\right)$,
\begin{align}\label{equ:LM1_2}
& G_i^{\top}\bar{C} \bar{A}^t \bar{x} + G_i^{\top} \left(\bar{C} \sum_{k=0}^{t-1} \bar{A}^k \bar{B}_v + \bar{D}_v\right) v \nonumber \\
&= G_i^{\top}\bar{C} \bar{A}^t \bar{x}^*(v) + G_i^{\top} \left(\bar{C} \sum_{k=0}^{t-1} \bar{A}^k \bar{B}_v + \bar{D}_v\right) v \nonumber \\
&\quad\quad + G_i^{\top}\bar{C} \bar{A}^t \left(\bar{x}-\bar{x}^*(v)\right) \nonumber \\[4pt]
&\le G_i^{\top} \left(\bar{C}(I-\bar{A})^{-1} \bar{B}_v + \bar{D}_v \right) v + \|G_i^{\top}\bar{C} \bar{A}^t\|\, \|\bar{x}-\bar{x}^*(v)\| \nonumber \\
&\le G_i^{\top} \left(\bar{C}(I-\bar{A})^{-1} \bar{B}_v + \bar{D}_v \right) v + \bar{\varepsilon}_i \nonumber \\
&\le g_i - \sqrt{2G_i^{\top} \Sigma_y(t) G_i}\, \text{erf}^{-1} \big(2\beta_i - 1\big),
\end{align}
for all $t \in \mathbb{Z}_{\ge 0}$ and $i = 1,...,n_g$. Note that since $\bar{A}$ is Schur (see Assumption~1), $\sup_{t \ge 0} \|G_i^{\top}\bar{C} \bar{A}^t\|$ is finite. According to the definition of $\tilde{\mathcal{O}}_{\infty}$ in \eqref{equ:O_i2} and \eqref{equ:O_all2}, the above \eqref{equ:LM1_2} implies $\mathcal{B}_{\varepsilon}\left(\bar{x}^*(v)\right) \times \{v\} \subset \tilde{\mathcal{O}}_{\infty}$ for any $v \in \bigcap_{i=1}^{n_g} \tilde{\Omega}^i$. This completes the proof. $\blacksquare$

We now show the almost-sure finite-time convergence property of the modified reference $v(t)$ to constant, steady-state constraint-admissible commands in the following proposition. We remark that this almost-sure finite-time convergence result is a stronger convergence result than the convergence in probability result established in Theorem~7 of \cite{kalabic2019reference} for the stochastic RG.

{\bf Proposition 3:} Suppose (i) $\left(x_p(0),x_u(0),v(0)\right) \in \tilde{\mathcal{O}}_{\infty}$, (ii) $\Sigma_x^{\infty}$ is nonsingular, and (iii) there exists $t_s \in \mathbb{Z}_{\ge 0}$ such that $r(t) = r_s \in \bigcap_{i=1}^{n_g} \tilde{\Omega}^i$ for all $t \in \mathbb{Z}_{\ge t_s}$. Then, the modified reference $v(t)$ almost surely (a.s.) converges to $r_s$ in finite time, i.e.,
\begin{equation}\label{equ:P3}
\mathbb{P} \Big\{\exists t_f \in \mathbb{Z}_{\ge t_s} \text{ such that } v(t) = r_s, \forall t \in \mathbb{Z}_{\ge t_f} \Big\} = 1.
\end{equation}

{\bf Proof:} Firstly, the constraint \eqref{equ:opt_33} ensures that any realization of the sequence $\left\{\|v(t)-r_s\|_R\right\}_{t=t_s}^{\infty}$ generated by the CSRG must be nonincreasing. Since $\|v(t)-r_s\|_R$ is also bounded from below by $0$, any realization of $\left\{\|v(t)-r_s\|_R\right\}_{t=t_s}^{\infty}$ must converge to some $\eta^* \in \mathbb{R}_{\ge 0}$. More specifically, with \eqref{equ:opt_33}, whenever $v(t)$ differs from its previous value $v(t-1)$, it must hold that either $\|v(t)-r_s\|_R \le \|v(t-1)-r_s\|_R - \delta$, with $\delta$ being a positive constant, or $\|v(t)-r_s\|_R = 0$. This ensures that any realization of $\left\{\|v(t)-r_s\|_R\right\}_{t=t_s}^{\infty}$ must converge to its corresponding $\eta^*$ through at most a finite number of jumps and the sequence $\left\{v(t)\right\}_{t=t_s}^{\infty}$ converges to some $v^* \in \mathbb{R}^{n_v}$ after these jumps. Note that due to the stochastic nature of the system \eqref{equ:system_cpt}, the $\eta^*$ and $v^*$ are random variables. In what follows we show that with probability $1$, $\eta^* = 0$ and $v^* = r_s$.

Assume the opposite, i.e., $\left\{v(t)\right\}_{t=t_s}^{\infty}$ converges to some $v^* \neq r_s$. In the above we have shown that $\left\{v(t)\right\}_{t=t_s}^{\infty}$ must reach $v^*$ after a finite number of jumps, which also implies that $\left\{v(t)\right\}_{t=t_s}^{\infty}$ reaches $v^*$ in finite time. Denote the time instant when $\left\{v(t)\right\}_{t=t_s}^{\infty}$ reaches $v^*$ as $t^* \in \mathbb{Z}_{\ge t_s}$. In particular, $v(t) = v^* \neq r_s$ for all $t \in \mathbb{Z}_{\ge t^*}$. 

In this case, the constraint \eqref{equ:opt_33} would never be satisfied over $t \in \mathbb{Z}_{\ge t^*}$, and according to the CSRG algorithm \eqref{equ:opt_2}, $x_u(t) = \bar{x}_u(t) = A_p x_p(t-1) + A_u x_u(t-1) + D_v v(t-1)$ for all $t \in \mathbb{Z}_{\ge t^*}$. This implies that the sequence $\left\{\bar{x}(t)\right\}_{t=t^*}^{\infty} =  \left\{[x_p^{\top}(t),x_u^{\top}(t)]^{\top}\right\}_{t=t^*}^{\infty}$ would be a Gaussian Markov process generated by the recursion \eqref{equ:system_cpt_1} with $v(t) \equiv v^*$. Then, because $\Sigma_x^{\infty}$ is assumed to be nonsingular, the Strong Law of
Large Numbers for Markov chains \citep{meyn1991asymptotic} says that the following,
\begin{equation}\label{equ:ergodicthm_1}
\lim_{N \to \infty} \frac{1}{N} \sum_{t = t^*}^{t^*+N-1} f\big(\bar{x}(t)\big) \,\,\aseq\,\, \int_{\mathbb{R}^{n_{xp}+n_{xu}}} f \,\text{d}\mu^*,
\end{equation}
would hold for any positive Borel function $f$ on $\mathbb{R}^{n_{xp}+n_{xu}}$ and almost every initial condition $\bar{x}(t^*) \in \mathbb{R}^{n_{xp}+n_{xu}}$, where $\mu^*$ is the Gaussian measure with mean $\bar{x}^*(v^*)$ and covariance $\Sigma_x^{\infty}$ \citep{bogachev1998gaussian}.

Because $v^*$ is generated by the CSRG, it must hold that $v^* \in {\rm proj}_v (\tilde{\mathcal{O}}_{\infty}) \subset \bigcap_{i=1}^{n_g} \tilde{\Omega}^i$. Then, consider the $f$ in \eqref{equ:ergodicthm_1} as the indicator function of the open ball $\mathcal{B}_{\varepsilon/2}(\bar{x}^*(v^*)) \subset \mathbb{R}^{n_{xp}+n_{xu}}$, i.e., $f = \mathbb{I}_{\mathcal{B}_{\varepsilon/2}(\bar{x}^*(v^*))}$, with $\varepsilon$ defined as in Lemma~1, and obtain
\begin{align}\label{equ:ergodicthm_2}
& \lim_{N \to \infty} \frac{1}{N} \sum_{t=t^*}^{t^*+N-1} \mathbb{I}_{\mathcal{B}_{\varepsilon/2}(\bar{x}^*(v^*))} \big(\bar{x}(t)\big) \\ 
& \aseq\,\, \int_{\mathbb{R}^{n_{xp}+n_{xu}}} \mathbb{I}_{\mathcal{B}_{\varepsilon/2}(\bar{x}^*(v^*))} \,\text{d}\mu^* = \mu^* \big(\mathcal{B}_{\varepsilon/2}(\bar{x}^*(v^*))\big) > 0, \nonumber 
\end{align}
which implies the existence of $N' \in \mathbb{Z}_{\ge 0}$ such that $\sum_{t=t^*}^{t^*+N'-1} \mathbb{I}_{\mathcal{B}_{\varepsilon/2}(\bar{x}^*(v^*))} \big(\bar{x}(t)\big) > 0$. And this in turn implies the existence of $t' \in \mathbb{Z}_{[t^*,t^*+N'-1]} \subset \mathbb{Z}_{\ge t^*}$ such that $\bar{x}(t') = [x_p^{\top}(t'),\bar{x}_u^{\top}(t')]^{\top} \in \mathcal{B}_{\varepsilon/2}(\bar{x}^*(v^*))$. Then, according to Lemma~1, we have $\left(x_p(t'),\bar{x}_u(t'),v^* \right) \in \mathcal{B}_{\varepsilon/2}(\bar{x}^*(v^*)) \times \{v^*\} \subset \tilde{\mathcal{O}}_{\infty}$. 

Now let $\xi_i = \sup_{t \ge 0} \|G_i^{\top} (\bar{C} \sum_{k=0}^{t-1} \bar{A}^k \bar{B}_v + \bar{D}_v)\|$, which is finite because $\bar{A}$ is Schur, and consider $\Delta v$ satisfying $\|\Delta v\| \le \min \big\{\bar{\varepsilon}_i/(2\xi_i) : i=1,...,n_g \big\}$, with $\bar{\varepsilon}_i>0$ defined as in Proposition~1. Referring to \eqref{equ:LM1_2} in the proof of Lemma~1, we obtain that
\begin{align}\label{equ:P3_1}
& G_i^{\top}\bar{C} \bar{A}^t \bar{x}(t') + G_i^{\top} \left(\bar{C} \sum_{k=0}^{t-1} \bar{A}^k \bar{B}_v + \bar{D}_v\right) (v^* + \Delta v) \nonumber \\
&= G_i^{\top}\bar{C} \bar{A}^t \bar{x}^*(v^*) + G_i^{\top} \left(\bar{C} \sum_{k=0}^{t-1} \bar{A}^k \bar{B}_v + \bar{D}_v\right) v^* \nonumber \\
&+ G_i^{\top}\bar{C} \bar{A}^t \left(\bar{x}(t')-\bar{x}^*(v^*)\right) + G_i^{\top} \left(\bar{C} \sum_{k=0}^{t-1} \bar{A}^k \bar{B}_v + \bar{D}_v\right) \Delta v \nonumber \\[4pt]
&\le G_i^{\top} \left(\bar{C}(I-\bar{A})^{-1} \bar{B}_v + \bar{D}_v \right) v^* \nonumber \\
&\quad + \|G_i^{\top}\bar{C} \bar{A}^t\|\, \|\bar{x}(t')-\bar{x}^*(v)\| + \xi_i\, \|\Delta v\|, \nonumber \\[2pt]
&\le G_i^{\top} \left(\bar{C}(I-\bar{A})^{-1} \bar{B}_v + \bar{D}_v \right) v^* + \bar{\varepsilon}_i \nonumber \\[2pt]
&\le g_i - \sqrt{2G_i^{\top} \Sigma_y(t) G_i}\, \text{erf}^{-1} \big(2\beta_i - 1\big),
\end{align}
for all $t \in \mathbb{Z}_{\ge 0}$.

Furthermore, since $v^*$ and $r_s$ both belong to the convex set $\bigcap_{i=1}^{n_g} \tilde{\Omega}^i$, any $v^* + \Delta v$ that lies on the line segment connecting $v^*$ and $r_s$ also belongs to $\bigcap_{i=1}^{n_g} \tilde{\Omega}^i$. For such a case, $\Delta v$ can be written as $\Delta v = \lambda (r_s - v^*)$ for some $\lambda \in [0,1]$. Together with \eqref{equ:P3_1} and according to the definition of $\tilde{\mathcal{O}}_{\infty}$ in \eqref{equ:O_i2} and \eqref{equ:O_all2}, it holds that for any $\Delta v = \lambda (r_s - v^*)$ with $0 \le \lambda \le \bar{\lambda} = \min\big\{\min \big\{\bar{\varepsilon}_i/(2\xi_i\|r_s-v^*\|) : i=1,...,n_g \big\},1\big\}$, $\left(x_p(t'),\bar{x}_u(t'),v^* + \Delta v\right) \in \tilde{\mathcal{O}}_{\infty}$.

Recall that $\delta>0$ is a sufficiently small constant. Now consider $\delta \le \min \big\{\bar{\varepsilon}_i/(2\xi_i c): i = 1,...,n_g \big\}$, where $c>0$ is a constant such that $\|\cdot\| \le c\, \|\cdot\|_R$ (according to the equivalence of norms on $\mathbb{R}^{n_v}$). Then, for $\Delta v = \bar{\lambda} (r_s - v^*)$, we have
\begin{align}\label{equ:P3_2}
    & \|v^* - r_s\|_R - \|v^* + \Delta v - r_s\|_R \nonumber \\
    &= \|v^* - r_s\|_R - \|v^*+ \bar{\lambda} (r_s - v^*) - r_s\|_R \nonumber \\
    &= \|v^* - r_s\|_R - (1-\bar{\lambda})\|v^* - r_s\|_R = \bar{\lambda}\, \|v^* - r_s\|_R \nonumber \\
    &= \min\left\{\min \left\{ \frac{\bar{\varepsilon}_i \|v^* - r_s\|_R }{2\xi_i\|v^*-r_s\|} : i=1,...,n_g \right\},\|v^* - r_s\|_R \right\} \nonumber \\
    &\ge \min\left\{\min \left\{ \frac{\bar{\varepsilon}_i }{2\xi_i c} : i=1,...,n_g \right\},\|v^* - r_s\|_R \right\} \nonumber \\
    &\ge \min\left\{\delta,\|v^* - r_s\|_R\right\},
\end{align}
which can be equivalently written as $\|v^* + \Delta v - r_s\|_R^2 \le \left(\max\left\{\|v^* - r_s\|_R - \delta,0\right\}\right)^2$.

Therefore, we have shown that the pair $\left(\bar{x}_u(t'),v^* + \Delta v\right) = \left(\bar{x}_u(t'),v^* + \bar{\lambda} (r_s - v^*)\right)$ is a feasible solution to the optimization problem \eqref{equ:opt_3}. Together with the fact that $\left(x_p(t'),\bar{x}_u(t'),v^* \right) \in \tilde{\mathcal{O}}_{\infty}$ shown above and according to the CSRG algorithm \eqref{equ:opt_2}, we should have $\left(x_u(t'),v(t')\right) = \left(\hat{x}_u(t'),\hat{v}(t')\right)$ at $t' \in \mathbb{Z}_{\ge t^*}$, with $\left(\hat{x}_u(t'),\hat{v}(t')\right)$ being the optimal solution to \eqref{equ:opt_3}. In particular, the constraint \eqref{equ:opt_33} ensures $\hat{v}(t') \neq v^*$ (note that we have assumed $v^* \neq r_s$).

This contradicts our assumption that $\left\{v(t)\right\}_{t=t_s}^{\infty}$ converges to some $v^* \neq r_s$. Since at the beginning of the proof we have shown that any realization of the sequence $\left\{v(t)\right\}_{t=t_s}^{\infty}$ must converge to some point, such a contradiction says that the converged point must be $v^* = r_s$.

To sum up, we have shown that the sequence $\left\{v(t)\right\}_{t=t_s}^{\infty}$ generated by the CSRG almost surely converges to $r_s$ after a finite number of jumps, which also implies the convergence is in finite time. Note that the ``almost surely'' comes from the almost sure equality in \eqref{equ:ergodicthm_1} and \eqref{equ:ergodicthm_2}. This result can also be explicitly expressed as \eqref{equ:P3}. $\blacksquare$

On the basis of the convergence result of the modified reference $v(t)$ to constant, steady-state constraint-admissible commanded reference $r_s$ in Proposition~3, we now discuss the stability property of the commanded set-point $\bar{x}^*(r_s)$. In our considered stochastic setting, the stability of $\bar{x}^*(r_s)$ is characterized by the following Proposition~4 and Remark~1.

{\bf Proposition 4:} Let $t_f \in \mathbb{Z}_{\ge t_s}$ denote the time instant such that $v(t) = r_s$ for all $t \in \mathbb{Z}_{\ge t_f}$. Then, for $t \in \mathbb{Z}_{\ge t_f}$, the difference between $\bar{x}(t)$ and $\bar{x}^*(r_s)$ is exponentially bounded in mean square \citep{tarn1976observers}. In particular, we have
\begin{align}\label{equ:P4}
& \mathbb{E} \left\{\left\|\bar{x}(t) - \bar{x}^*(r_s)\right\|^2 \big|\bar{x}(t_f)\right\} \le \nonumber \\
&\quad\quad \frac{\mu}{\alpha} + (1-\alpha)^{t-t_f} \|\bar{x}(t_f) - \bar{x}^*(r_s)\|^2,
\end{align}
for some constants $\mu>0$ and $\alpha \in (0,1]$.

{\bf Proof:} Let $t \in \mathbb{Z}_{\ge t_f}$ be arbitrary. Define $\zeta(t) = \bar{x}(t) - \bar{x}^*(r_s)$ and consider the function $V\left(\zeta(t)\right) = \|\zeta(t)\|_P^2 = J\left(x_u(t),r_s,x_p(t),r_s\right)$. If $(x_p(t+1),\bar{x}_u(t+1),v(t) = r_s) \in \tilde{\mathcal{O}}_{\infty}$, then \eqref{equ:opt_2} yields that $\left(x_u(t+1),v(t+1)\right) = \left(\hat{x}_u(t+1),\hat{v}(t+1)\right)$ with $\left(\hat{x}_u(t+1),\hat{v}(t+1)\right)$ determined by \eqref{equ:opt_3}. In this case, both $\left(\bar{x}_u(t+1),r_s\right)$ and $\left(\hat{x}_u(t+1),\hat{v}(t+1)\right)$ are feasible solutions to \eqref{equ:opt_3}, with $\left(\hat{x}_u(t+1),\hat{v}(t+1)\right)$ being the optimal one. Indeed, the constraint \eqref{equ:opt_33} also ensures $\hat{v}(t+1) = r_s$. As a result, we must have $V\left(\zeta(t+1)\right) = J\left(\hat{x}_u(t+1),r_s,x_p(t+1),r_s\right) \le J(\bar{x}_u(t+1),r_s,x_p(t+1),r_s)$. If $\left(x_p(t+1),\bar{x}_u(t+1),v(t) = r_s\right) \notin \tilde{\mathcal{O}}_{\infty}$, then \eqref{equ:opt_2} yields that $\left(x_u(t+1),v(t+1)\right) = \left(\bar{x}_u(t+1),v(t) = r_s\right)$, and in turn, $V\left(\zeta(t+1)\right) = J\left(x_u(t+1),r_s,x_p(t+1),r_s\right) = J\left(\bar{x}_u(t+1),r_s,x_p(t+1),r_s\right)$. 

For given $\bar{x}(t) = [x_p^{\top}(t), x_u^{\top}(t)]^{\top}$, we have
\begin{align}\label{equ:P41}
& \mathbb{E}\left\{J\left(\bar{x}_u(t+1),r_s,x_p(t+1),r_s\right) \big|\bar{x}(t)\right\} \nonumber \\
&= \mathbb{E}\left\{ \left\|\bar{A} \bar{x}(t) + \bar{B}_v r_s + \bar{B}_w w(t) - (I-\bar{A})^{-1} \bar{B}_v r_s\right\|_{P}^2 \Big|\bar{x}(t) \right\} \nonumber \\
&= \mathbb{E}\left\{ \left\| \bar{A} \zeta(t) + \bar{B}_w w(t) \right\|_{P}^2 \Big|\zeta(t) \right\} \nonumber \\
&= \mathbb{E}\Big\{\zeta(t)^{\top}\left(\bar{A}^{\top}P\bar{A}\right)\zeta(t) + 2w(t)^{\top}\left(\bar{B}_w^{\top}P\bar{A}\right)\zeta(t) \nonumber \\
&\quad\quad\quad\quad + w(t)^{\top}\left(\bar{B}_w^{\top}P\bar{B}_w\right)w(t) \Big|\zeta(t) \Big\} \nonumber \\
&= \zeta(t)^{\top}\left(P-Q\right)\zeta(t) + \mathbb{E}\left\{w(t)^{\top}\left(\bar{B}_w^{\top}P\bar{B}_w\right)w(t)\right\} \nonumber \\
&= V\left(\zeta(t)\right) - \|\zeta(t)\|_Q^2 + \text{trace}\left(W\left(\bar{B}_w^{\top}P\bar{B}_w\right)\right).
\end{align}
Then, using the pointwise inequality $V\left(\zeta(t+1)\right) \le J\left(\bar{x}_u(t+1),r_s,x_p(t+1),r_s\right)$ shown above and the equivalence of norms on $\mathbb{R}^{n_{xp}+n_{xu}}$, we obtain
\begin{align}\label{equ:P42}
& \mathbb{E}\left\{V\left(\zeta(t+1)\right)\big|\zeta(t)\right\} \nonumber \\
&\le \mathbb{E}\left\{J\left(\bar{x}_u(t+1),r_s,x_p(t+1),r_s\right) \big|\bar{x}(t)\right\} \nonumber \\
&\le \mu + (1-\alpha) V\left(\zeta(t)\right),
\end{align}
where $\mu = \text{trace}\left(W\left(\bar{B}_w^{\top}P\bar{B}_w\right)\right)$ and $\alpha \in (0,1]$ is such that $\sqrt{\alpha}\,\|\cdot\|_P \le \|\cdot\|_Q$.

Since the $t \in \mathbb{Z}_{\ge t_f}$ is arbitrary, \eqref{equ:P42} also yields
\begin{align}\label{equ:P43}
& \mathbb{E}\left\{V\left(\zeta(t+2)\right)\big|\zeta(t)\right\} \nonumber \\
&= \mathbb{E}\left\{\mathbb{E}\left\{V\left(\zeta(t+2)\right)\big|\zeta(t+1)\right\}\Big|\zeta(t)\right\} \nonumber \\
&\le \mathbb{E}\left\{\mu + (1-\alpha) V\left(\zeta(t+1)\right) \big|\zeta(t)\right\} \nonumber \\
&= \mu + (1-\alpha)\, \mathbb{E}\left\{V\left(\zeta(t+1)\right)\big|\zeta(t)\right\} \nonumber \\
&\le \mu + (1-\alpha) \big(\mu + (1-\alpha) V(\zeta(t))\big) \nonumber \\
&= \left(\sum_{i = 0}^{1} (1-\alpha)^i\right)\mu + (1-\alpha)^2 V\left(\zeta(t)\right).
\end{align}
Continuing this way, we obtain
\begin{align}\label{equ:P44}
& \mathbb{E}\left\{V\left(\zeta(t_f+n)\right)\big|\zeta(t_f)\right\} \nonumber \\
&= \left(\sum_{i = 0}^{n-1} (1-\alpha)^i\right)\mu + (1-\alpha)^n V\left(\zeta(t_f)\right) \nonumber \\
&\le \frac{\mu}{\alpha} + (1-\alpha)^n V\left(\zeta(t_f)\right).
\end{align}
Then, \eqref{equ:P4} follows from \eqref{equ:P44} and the equivalence of norms on $\mathbb{R}^{n_{xp}+n_{xu}}$. $\blacksquare$

{\bf Remark 1:} Note first that the existence of the finite time $t_f \in \mathbb{Z}_{\ge t_s}$ in Proposition~4 is almost surely guaranteed (see Proposition~3). The difference between $\bar{x}(t)$ and the commanded set-point $\bar{x}^*(r_s)$ being exponentially bounded in mean square as in \eqref{equ:P4} also yields that $\bar{x}^*(r_s)$ is asymptotically stable in mean square in the large \citep{tarn1976observers}, since as $t \to \infty$,
\begin{equation}
  \mathbb{E}\left\{\lim_{t \to \infty} \left\|\bar{x}(t) - \bar{x}^*(r_s)\right\|^2 \right\} \le \frac{\mu}{\alpha}.
\end{equation}
Note that we have dropped the condition on $\bar{x}(t_f)$ from the expectation since the right-hand side does not depend on $\bar{x}(t_f)$ \citep{tarn1976observers}.

\subsection{Joint chance constraints}

The CSRG scheme introduced in Section~\ref{sec:3} is designed to enforce the set of individual chance constraints \eqref{equ:chance_constraint}. However, one may sometimes be interested in enforcing chance constraints in the following form,
\begin{equation}\label{equ:joint_constraint}
\mathbb{P} \left\{G_i^{\top} y(t) \le g_i,\,  i = 1,...,n_g \right\} \ge \beta,
\end{equation}
with $\beta \in (0.5,1)$, called a joint chance constraint. In this subsection, we extend the CSRG scheme to treat such joint chance constraints.

In principle, one could define a maximal output admissible set $\mathcal{O}_{\infty}$ similar to the one defined in \eqref{equ:O_all} and \eqref{equ:O_i} but corresponding to the joint chance constraint \eqref{equ:joint_constraint} as follows,
\begin{align}\label{equ:O_joint}
& \mathcal{O}_{\infty} = \big\{(x_p,x_u,v) \in \mathbb{R}^{n_{xp}} \!\times\! \mathbb{R}^{n_{xu}} \!\times\! \mathbb{R}^{n_v}:\, \text{if } x_p(0) = x_p, \nonumber \\
& x_u(0) = x_u, v(t) \equiv v, \text{ and } \{w(t)\}_{t \in \mathbb{Z}_{\ge 0}} \sim \mathcal{N}(0,W), \nonumber \\
& \text{then } \mathbb{P} \big\{G_i^{\top} y(t) \le g_i,\,  i = 1,...,n_g \big\} \ge \beta \text{ for all } t \in \mathbb{Z}_{\ge 0} \big\}.
\end{align}
Then, with this new $\mathcal{O}_{\infty}$ set, one could formulate the CSRG algorithms similarly as before.

However, the exact construction of the above $\mathcal{O}_{\infty}$ set requires evaluation of $\mathbb{P} \big\{G_i^{\top} y(t) \le g_i,\,  i = 1,...,n_g \big\}$, which involves integration of the density function of a multivariate normal distribution over a polyhedral set and is in general computationally challenging \citep{khachiyan1989problem}. Furthermore, this $\mathcal{O}_{\infty}$ set cannot be characterized by a collection of linear inequalities as in \eqref{equ:O_i2}, which could also cause the online problems \eqref{equ:opt_3} and \eqref{equ:opt_5} to be difficult to solve.

Therefore, in what follows we pursue inner approximations of the above $\mathcal{O}_{\infty}$ set that are easier to compute and will use such approximations to formulate our CSRG algorithms instead of directly using $\mathcal{O}_{\infty}$. In particular, we consider the following two approximation approaches:

\subsubsection{Risk allocation}

Note the left-hand side of \eqref{equ:joint_constraint} can be lower bounded as
\begin{align}\label{equ:Boole}
& \mathbb{P} \left\{G_i^{\top} y(t) \le g_i,\,  i = 1,...,n_g \right\} = \mathbb{P} \left\{ \bigcap_{i=1}^{n_g} \left(G_i^{\top} y(t) \le g_i\right) \right\} \nonumber \\
&\!\! = 1 - \mathbb{P} \left\{ \bigcup_{i=1}^{n_g} \left(G_i^{\top} y(t) > g_i\right) \right\} \ge 1 - \sum_{i=1}^{n_g} \mathbb{P} \left\{G_i^{\top} y(t) > g_i \right\} \nonumber \\
&\!\! = 1 - \sum_{i=1}^{n_g} \left(1- \mathbb{P} \left\{ G_i^{\top} y(t) \le g_i \right\} \right)  \nonumber \\
&\!\! = \sum_{i=1}^{n_g} \mathbb{P} \left\{ G_i^{\top} y(t) \le g_i \right\} - (n_g - 1),
\end{align}
where we have used Boole's inequality in the second line. Then, it can be easily seen that one can enforce the joint chance constraint \eqref{equ:joint_constraint} through enforcing the set of individual chance constraints \eqref{equ:chance_constraint} with some $\beta_i$, $i = 1,...,n_g$, satisfying 
\begin{equation}\label{equ:risk_allocation}
\sum_{i=1}^{n_g} \beta_i \ge \beta + (n_g - 1).
\end{equation}
The parameters $\beta_i$, $i = 1,...,n_g$, could be treated as optimization variables to reduce conservativeness in the approximation \citep{blackmore2009convex,paulson2020stochastic}. This way, the constraint functions in \eqref{equ:O_i2} characterizing the $\mathcal{O}_{\infty}^i$ set would be nonlinear functions of the variables $(x_p,x_u,v,\beta_i)$. To render linear constraints so as to simplify both the offline construction of $\mathcal{O}_{\infty}^i$ and the online problems \eqref{equ:opt_3} and \eqref{equ:opt_5}, an alternative way is to pre-specify the values of $\beta_i$, $i = 1,...,n_g$, such that the condition \eqref{equ:risk_allocation} is satisfied. A typical choice as in \cite{nemirovski2007convex} is given by
\begin{equation}\label{equ:risk_allocation_2}
\beta_i = \frac{\beta + (n_g - 1)}{n_g}, \quad i = 1, \cdots, n_g.
\end{equation}
To sum up, to treat the joint chance constraint \eqref{equ:joint_constraint}, the CSRG algorithms are formulated as in Section~\ref{sec:3}, with the $\beta_i$, $i = 1,...,n_g$, determined according to \eqref{equ:risk_allocation_2}. Following \cite{blackmore2009convex,paulson2020stochastic}, this approach to treating joint chance constraints is referred to as the {\it risk allocation} approach.

\subsubsection{$\beta$-level confidence ellipsoid}

Another approach to guaranteeing satisfaction of the joint chance constraint \eqref{equ:joint_constraint} is to enforce the $\beta$-level confidence ellipsoid of $y(t)$ to be entirely contained in the constraint set. Specifically, the following set of conditions are enforced,
\begin{equation}\label{equ:ellipsoid_1}
G_i^{\top} \mathcal{P}(t) \le g_i, \quad i = 1, \cdots, n_g,
\end{equation}
in which $\mathcal{P}(t) = \big\{y \in \mathbb{R}^{n_y}: \left(y-\bar{y}(t)\right)^{\top} \left(\Sigma_y(t)\right)^{-1}$  $\left(y-\bar{y}(t)\right) \le F^{-1}(\beta,n_y) \big\}$ is the $\beta$-level confidence ellipsoid of $y(t) \sim \mathcal{N}\left(\bar{y}(t),\Sigma_y(t)\right)$, where $F(\cdot,n_y)$ denotes the cumulative distribution function of the $\chi^2$-distribution with $n_y$ degrees of freedom.

Note that \eqref{equ:ellipsoid_1} guarantees
\begin{align}\label{equ:ellipsoid_2}
& \mathbb{P} \left\{G_i^{\top} y(t) \le g_i,\,  i = 1,...,n_g \right\} \nonumber \\
&\ge \mathbb{P} \Big\{\big(G_i^{\top} y(t) \le g_i,\,  i = 1,...,n_g\big) \cap \big(y(t) \in \mathcal{P}(t)\big) \Big\} \nonumber \\
&= \mathbb{P} \big\{y(t) \in \mathcal{P}(t)\big\} = \beta,
\end{align}
where we have used \eqref{equ:ellipsoid_1} to derive the equality between the second and the third lines. Note also that the support function of the confidence ellipsoid $\mathcal{P}(t)$ is $h_{\mathcal{P}(t)}(y) = y^{\top}\bar{y}(t) + \sqrt{F^{-1}(\beta,n_y)\, y^{\top} \Sigma_y(t) y}$ \citep{kurzhanski2000ellipsoidal}. Then, according to Theorem~2.3 of \cite{kolmanovsky1998theory}, the conditions in \eqref{equ:ellipsoid_1} can be equivalently expressed as
\begin{align}\label{equ:ellipsoid_3}
G_i^{\top} \bar{y}(t) &=G_i^{\top} \bar{C} \bar{A}^t \begin{bmatrix} x_p(0) \\ x_u(0) \end{bmatrix} + G_i^{\top} \left(\bar{C} \sum_{k=0}^{t-1} \bar{A}^k \bar{B}_v + \bar{D}_v\right) v \nonumber \\
&\le g_i - \sqrt{F^{-1}\big(\beta,n_y\big)\, G_i^{\top} \Sigma_y(t) G_i}\, .
\end{align}
Therefore, to treat the joint chance constraint \eqref{equ:joint_constraint}, the CSRG algorithms are formulated similarly as in Section~\ref{sec:3}, but with the linear inequalities characterizing the $\mathcal{O}_{\infty}^i$ set, \eqref{equ:chance_constraint_erf}, replaced with the above \eqref{equ:ellipsoid_3}. This approach to treating joint chance constraints has been exploited in \cite{van2006stochastic,kalabic2019reference} and is referred to as the {\it confidence ellipsoid} approach.

\subsubsection{Comparison}

The risk allocation approach and the confidence ellipsoid approach both can guarantee the satisfaction of the joint chance constraint \eqref{equ:joint_constraint}, shown in \eqref{equ:Boole} and \eqref{equ:ellipsoid_2}, respectively. However, probability inequalities are exploited in \eqref{equ:Boole} and \eqref{equ:ellipsoid_2} to achieve this guarantee, which typically cause the feasible set to shrink and result in conservativeness. We now compare the relative conservativeness of the two approaches.

Recall that in the risk allocation approach, \eqref{equ:joint_constraint} is converted into a set of constraints on the mean $\bar{y}(t)$ as follows:
\begin{align}\label{equ:compare_RA}
& G_i^{\top} \bar{y}(t) \le \\
& g_i - \sqrt{2}\, \text{erf}^{-1} \left(2\frac{\beta + (n_g - 1)}{n_g} - 1\right) \sqrt{G_i^{\top} \Sigma_y(t) G_i}, \nonumber
\end{align}
for $i = 1,...,n_g$; while in the confidence ellipsoid approach, the following set of constraints on $\bar{y}(t)$ are enforced:
\begin{align}\label{equ:compare_CE}
& G_i^{\top} \bar{y}(t) \le g_i - \sqrt{F^{-1}\big(\beta,n_y\big)}\, \sqrt{G_i^{\top} \Sigma_y(t) G_i}\, ,
\end{align}
for $i = 1,...,n_g$.

It can be seen that \eqref{equ:compare_RA} and \eqref{equ:compare_CE} are both linear inequalities on $\bar{y}(t)$, with the same left-hand side $G_i^{\top} \bar{y}(t)$ and constant right-hand sides. Note that the right-hand sides are constants because the covariance $\Sigma_y(t)$, determined according to \eqref{equ:covariance_2}, is independent of the variables $\left(x_p(0),x_u(0),v\right)$. Based on such an observation, the relative conservativeness of the risk allocation approach and the confidence ellipsoid approach can be determined by comparing the right-hand sides of \eqref{equ:compare_RA} and \eqref{equ:compare_CE}. Specifically, in the case where \eqref{equ:compare_RA} has a greater right-hand side than \eqref{equ:compare_CE}, which implies that the feasible set characterized by \eqref{equ:compare_RA} is a superset of that characterized by \eqref{equ:compare_CE}, the risk allocation approach is less conservative than the confidence ellipsoid approach; and vice versa.

Furthermore, which one between the right-hand sides of \eqref{equ:compare_RA} and \eqref{equ:compare_CE} is greater can be determined by the sign of the following function,
\begin{align}\label{equ:compare_func}
& \Gamma(n_y,n_g,\beta) = \\
& \sqrt{2}\, \text{erf}^{-1} \left(2\, \frac{\beta + (n_g - 1)}{n_g} - 1\right) - \sqrt{F^{-1}\big(\beta,n_y\big)}\, , \nonumber
\end{align}
which depends only on the dimension of the output vector $n_y$, the number of constraints $n_g$ and the required confidence level of constraint satisfaction $\beta$, but not on time $t$, specific system dynamics \eqref{equ:system_cpt} or constraints $(G_i,g_i)$. In particular, the relationship between the sign of \eqref{equ:compare_func} and the relative conservativeness of the two approaches is characterized by the following proposition.

{\bf Proposition 5:} If in \eqref{equ:compare_func}, $\Gamma(n_y,n_g,\beta) < 0$, then the feasible set corresponding to the risk allocation approach is a strict superset of that corresponding to the confidence ellipsoid approach, i.e., the risk allocation approach is less conservative than the confidence ellipsoid approach; if $\Gamma(n_y,n_g,\beta) = 0$, then the two approaches are equally conservative; if $\Gamma(n_y,n_g,\beta) > 0$, then the risk allocation approach is more conservative than the confidence ellipsoid approach.

{\bf Proof:} It can be easily seen that if $\Gamma(n_y,n_g,\beta) < 0$, then for any $i = 1,...,n_g$ and $t \in \mathbb{Z}_{\ge 0}$, \eqref{equ:compare_RA} has a strictly greater right-hand side than \eqref{equ:compare_CE}, which implies that the feasible set characterized by \eqref{equ:compare_RA} is a strict superset of that characterized by \eqref{equ:compare_CE}, i.e., the risk allocation approach is less conservative than the confidence ellipsoid approach. The other two cases can be shown in a similar way. $\blacksquare$

{\bf Remark 2:} Although stated in the context of our chance-constrained CSRG, Proposition~5 represents a more general result on the relative conservativeness of the risk allocation approach with equal risks \eqref{equ:risk_allocation_2} and the confidence ellipsoid approach to treating the joint chance constraint \eqref{equ:joint_constraint}. This result can also be used in other chance-constrained control techniques, such as in stochastic model predictive control \citep{mesbah2016stochastic}.

\begin{figure}[ht!]
\begin{center}
\begin{picture}(200.0, 180.0)
\put(  0,  -2){\epsfig{file=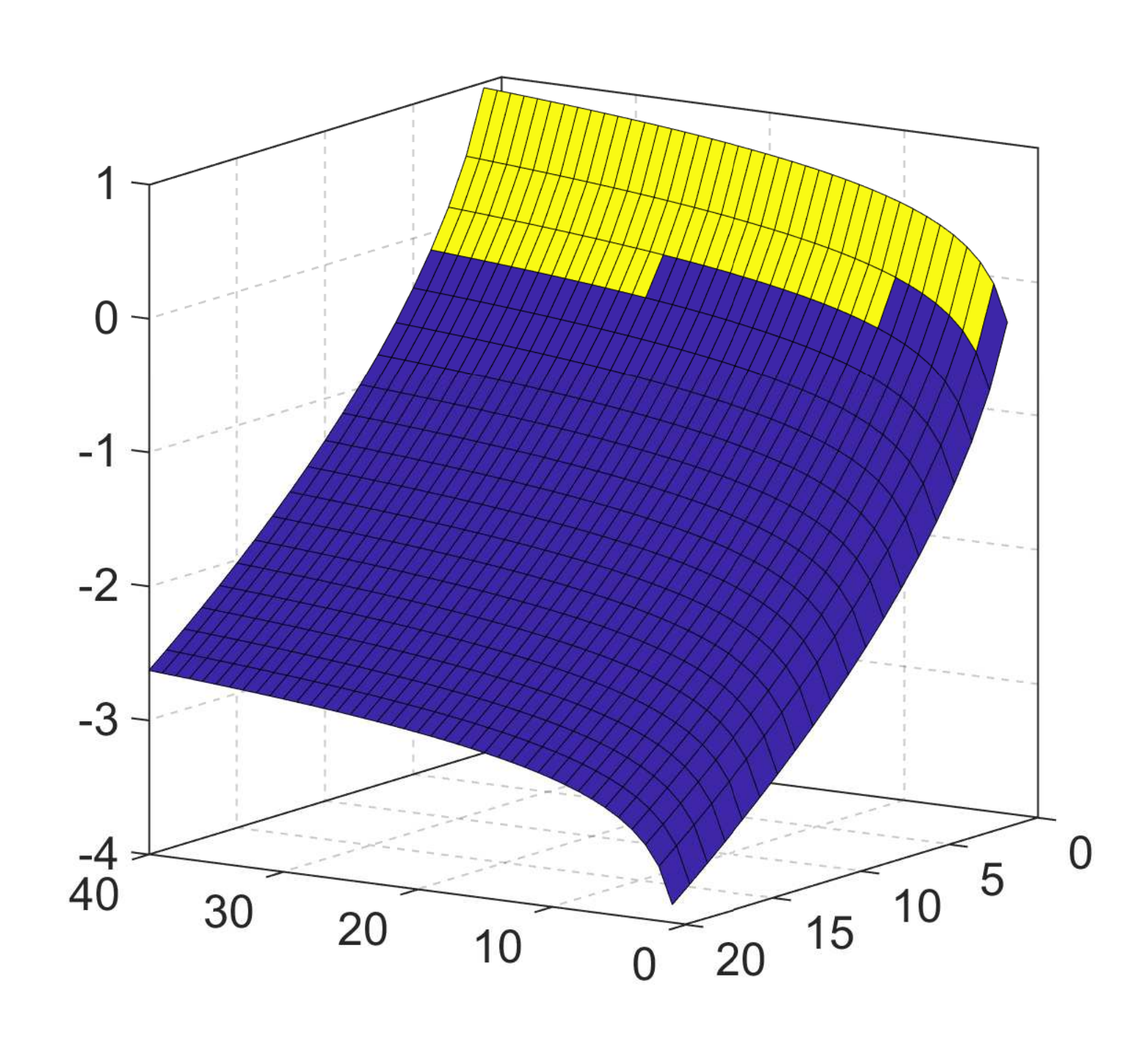,height=2.5in}}  
\put(  50,  4.4){$n_g$}
\put(  160,  8){$n_y$}
\put(  0,  88){$\Gamma$}
\end{picture}
\end{center}
      \caption{Graph of $\Gamma(n_y,n_g,\beta)$ for $\beta = 0.98$.}
      \label{fig:Gamma}
\end{figure}
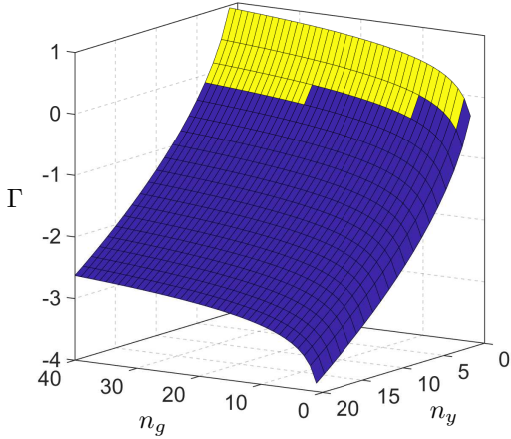

Fig.~\ref{fig:Gamma} shows the graph of the function $\Gamma(n_y,n_g,\beta)$ when $\beta$ is fixed at $0.98$. In Fig.~\ref{fig:Gamma}, the data points marked by blue (yellow) correspond to the cases where the risk allocation approach (the confidence ellipsoid approach) is less conservative. It can be seen that for cases with small $n_y$ and large $n_g$, the confidence ellipsoid approach is less conservative; the risk allocation approach is less conservative for all other cases. 

For a specific joint chance constrained problem with given $(n_y,n_g,\beta)$ values, one can refer to Proposition~5 to choose between the risk allocation approach and the confidence ellipsoid approach to achieve a less conservative CSRG design.

\section{CSRG Application to Aircraft Flight Control}\label{sec:5}

We now use numerical examples to illustrate the CSRG application to constrained control of aircraft. The aircraft models used in this section are taken from \cite{mcdonough2015controller,mcdonough2015integrator}, which are generated using the NASA generic transport model (GTM) \citep{cunningham2008practical} at the trim condition of altitude $h_0 = 800$ ft, airspeed $U_0 = 118.15$ ft/s ($70$ knots), sideslip angle $\beta_0 = 0$ rad, flight path angle $\gamma_0 = 0$ rad, and yaw rate $\dot{\psi}_0 = 0$ rad/s. 

\subsection{Constrained longitudinal flight control}

The aircraft longitudinal dynamics are represented by the following continuous-time linear model, 
\vspace{-6mm}\small 
\begin{align}\label{equ:longitudinal_sys}
& \begin{bmatrix} \Delta \dot{U} \\ \Delta \dot{\alpha} \\ \Delta \dot{q} \\ \Delta \dot{\theta} \end{bmatrix} = \underbrace{\begin{bmatrix} -0.0665 & -11.4608 & 0.1439 & -32.1740 \\
-0.0035 & -2.4714 & 0.9514 & 0 \\
-0.0090 & -43.9070 & -3.4738 & 0 \\
0 & 0 & 1 & 0 \end{bmatrix}}_{= A^{\text{lon}}} \begin{bmatrix} \Delta U \\ \Delta \alpha \\ \Delta q \\ \Delta \theta \end{bmatrix} \nonumber \\
&+ \underbrace{\begin{bmatrix} -0.0435 & 0.1424 \\
-0.0043 & -0.0001 \\
-0.7662 & 0.0192 \\
0 & 0 \end{bmatrix}}_{= B_u^{\text{lon}}} \begin{bmatrix} \Delta \delta_e \\ \Delta \delta_T \end{bmatrix} -  \underbrace{\begin{bmatrix} -0.0665 & -11.4608 \\
-0.0035 & -2.4714 \\
-0.0090 & -43.9070 \\
0 & 0 & \end{bmatrix}}_{= B_w^{\text{lon}}} \begin{bmatrix} w_x \\ w_z \end{bmatrix},
\end{align} 
\normalsize
where the states $x_p^{\text{lon}} = [\Delta U, \Delta \alpha, \Delta q, \Delta \theta]^{\top}$ represent the deviations in longitudinal airspeed (ft/s), angle of attack (rad), pitch rate (rad/s) and pitch angle (rad), respectively, the control inputs $u^{\text{lon}} = [\Delta \delta_e, \Delta \delta_T]^{\top}$ represent the deviations in elevator angle (rad) and thrust throttle percentage, and $w^{\text{lon}} = [w_x, w_z]^{\top}$ represents effects of model mismatch, atmospheric disturbances/turbulence, etc. in the longitudinal and vertical directions. The treatment of more general, dynamic models for atmospheric disturbances is out of scope of this paper but will be addressed in future work.

The continuous-time model \eqref{equ:longitudinal_sys} is converted to a discrete-time model using zero-order hold on the inputs with a sampling period of $\Delta T = 0.1$ s for the nominal controller and our CSRG design.

We consider a state-feedback controller with integral action for tracking commanded flight path angles $\Delta \gamma = -\Delta \alpha + \Delta \theta$. The controller takes the form of \eqref{equ:system_controller}, where
\begin{align}\label{equ:longitudinal_control_1}
& K_p^{\text{lon}} = \begin{bmatrix} -0.4735 & -37.7045 & 2.4948 & 46.3031 \\
   -2.4179 & 38.5827 & 0.2705 & -33.6410 \end{bmatrix}, \nonumber \\
& K_u^{\text{lon}} = \begin{bmatrix} 2.2715 \\ -6.1106 \end{bmatrix}, \quad\quad B_v^{\text{lon}} = \begin{bmatrix} 0 \\ 0 \end{bmatrix},  
\end{align}
with the integrator
\begin{equation}\label{equ:longitudinal_control_2}
x_u^{\text{lon}}(t+1) = \underbrace{\big[\, 0\; -1\;\; 0\;\; 1\, \big]}_{= A_p^{\text{lon}}} x_p^{\text{lon}} + x_u^{\text{lon}}(t) - \Delta \gamma_{\text{com}}(t).
\end{equation}
We note that this controller design is motivated by the longitudinal flight controllers proposed in \cite{mcdonough2015controller,mcdonough2015integrator}. In particular, the gains $K^{\text{lon}} = [K_p^{\text{lon}}, K_u^{\text{lon}}]$ are obtained by solving a discrete-time Linear-Quadratic-Regulator (LQR) problem with $\bar{A}_{\text{open}}^{\text{lon}} = \begin{bmatrix} A^{\text{lon}} & 0 \\ A_p^{\text{lon}} & 1 \end{bmatrix}$, $\bar{B}_u^{\text{lon}} = \begin{bmatrix} B_u^{\text{lon}} \\ 0 \end{bmatrix}$ and $Q^{\text{lon}} = \text{diag}(10, 10, 10, 10, 100)$, $R^{\text{lon}} = \text{diag}(10, 1)$.

\begin{figure}[ht!]
\begin{center}
\begin{picture}(240.0, 126.0)
\put(  0,  0){\epsfig{file=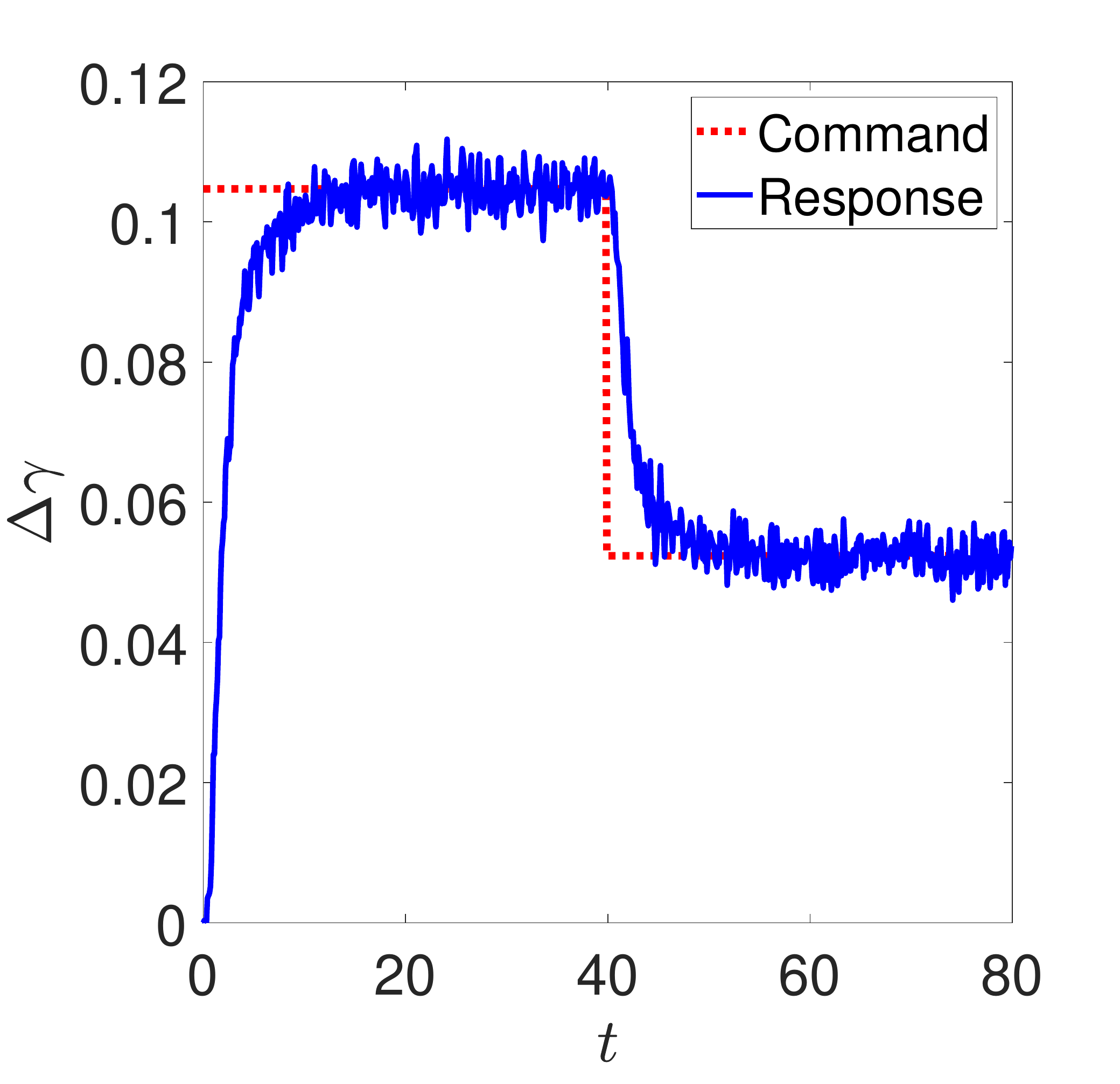,height=.24\textwidth}}
\put(  120,  0){\epsfig{file=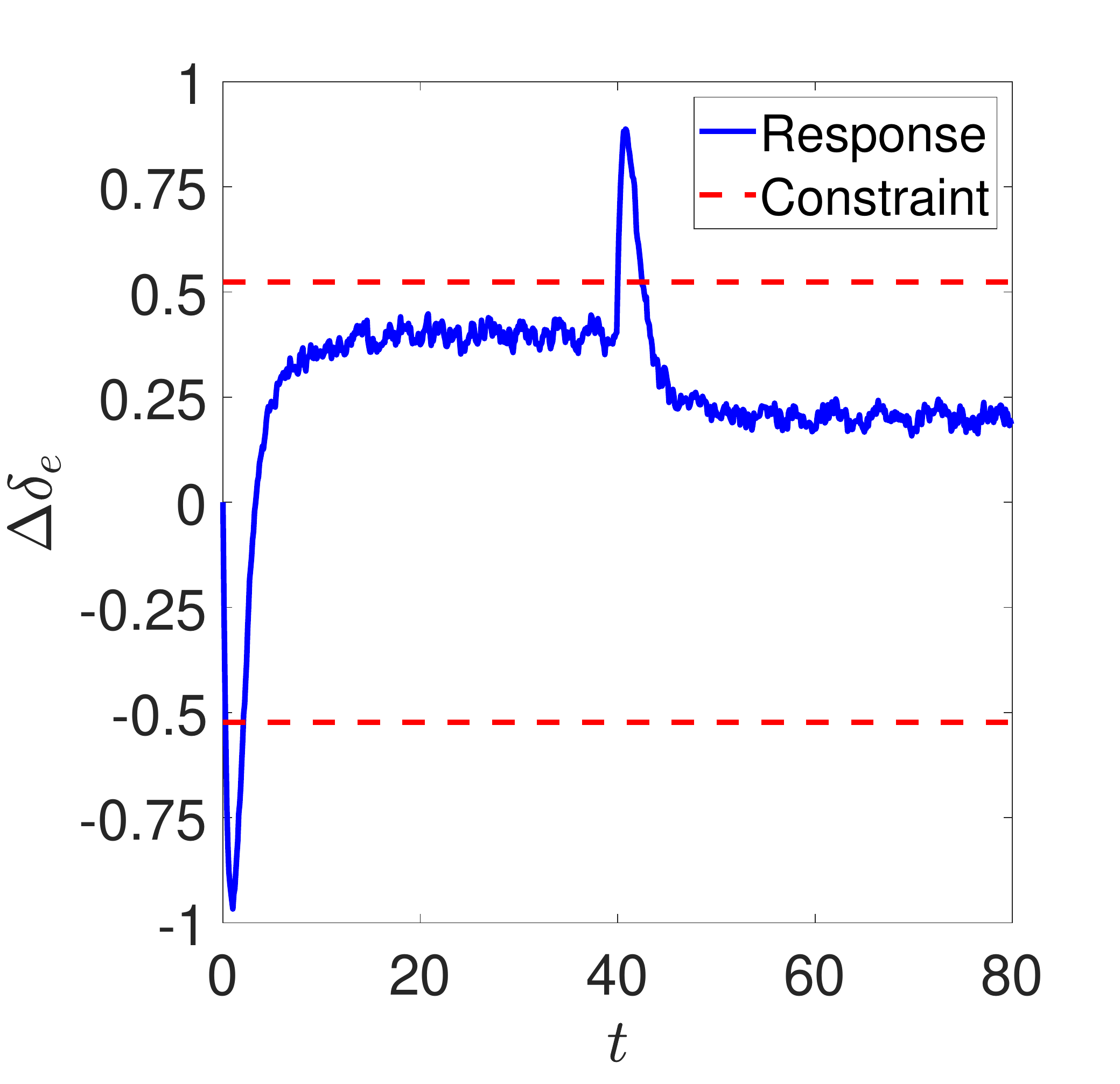,height=.24\textwidth}}
\small
\put( 100, 116){(a)}
\put( 220, 116){(b)}
\normalsize

\end{picture}
\end{center}
      \caption{Longitudinal flight control: Simulated trajectory of the nominal closed-loop system.}
      \label{fig:lon_1}
\end{figure}

We assume that the aircraft operation is subject to the following set of constraints on the states and control inputs,
\begin{align}\label{equ:longitudinal_constraints}
& -20 \le \Delta U(t) \le 20, &  -\frac{\pi}{32} \le \Delta \alpha(t) \le \frac{\pi}{24}, & \nonumber \\
& -\frac{\pi}{12} \le \Delta q(t) \le \frac{\pi}{12}, & -\frac{\pi}{6} \le \Delta \theta(t) \le \frac{\pi}{6}, & \\
& -\frac{\pi}{6} \le \Delta \delta_e(t) \le \frac{\pi}{6}, & -25 \le \Delta \delta_T(t) \le 25. & \nonumber 
\end{align}
The disturbance $w^{\text{lon}} = [w_x, w_z]^{\top}$ is modeled as a Gaussian noise. In particular, we assume the inputs $\{w^{\text{lon}}(t)\}_{t \in \mathbb{Z}_{\ge 0}}$ to satisfy Assumption~2 with
\begin{equation}\label{equ:longitudinal_disturbance}
\{w^{\text{lon}}(t)\}_{t \in \mathbb{Z}_{\ge 0}} \sim \mathcal{N} \left(0, \text{diag}(10^{-2},10^{-4})\right).
\end{equation}
Firstly, we simulate the response of the nominal closed-loop system (consisting of the plant \eqref{equ:longitudinal_sys} and the nominal controller \eqref{equ:longitudinal_control_1}-\eqref{equ:longitudinal_control_2}) starting from the initial condition $x_p^{\text{lon}}(0) = [0,0,0,0]^{\top}$ and $x_u^{\text{lon}}(0) = 0$ to track the commanded flight path angle profile $\Delta \gamma_{\text{com}}(t)$ shown by the red dotted curve in Fig.~\ref{fig:lon_1}(a). The trajectories of the actual flight path angle $\Delta \gamma(t) = -\Delta \alpha(t) + \Delta \theta(t)$ and the elevator angle $\Delta \delta_e(t)$ are plotted in Fig.~\ref{fig:lon_1}. It can be observed from Fig.~\ref{fig:lon_1}(b) that the transient responses of $\Delta \delta_e(t)$ to the step changes in $\Delta \gamma_{\text{com}}(t)$ significantly violate the constraints $-\frac{\pi}{6} \le \Delta \delta_e(t) \le \frac{\pi}{6}$.

We now apply the proposed CSRG scheme to enforce the constraints \eqref{equ:longitudinal_constraints}. Due to the presence of the stochastic disturbances \eqref{equ:longitudinal_disturbance}, we enforce the constraints \eqref{equ:longitudinal_constraints} probabilistically. In this example, we enforce each of them as an individual chance constraint in the form of \eqref{equ:chance_constraint}, with $\beta_i = 0.99$ for all $i = 1,...,12$. For the CSRG online optimization problem, we choose $P$ as the positive-definite solution to \eqref{equ:P_matrix} with $Q = \text{diag}(1, 1, 1, 1, 1)$ and $R = 10^4$ in the cost function \eqref{equ:cost}. We remark that a large $R$ can increase the convergence speed of the modified reference $v(t)$ to the commanded value $r(t)$. Also, we choose $\delta = 10^{-6}$ in the constraint \eqref{equ:opt_33}.

\begin{figure}[ht!]
\begin{center}
\begin{picture}(240.0, 126.0)
\put(  0,  0){\epsfig{file=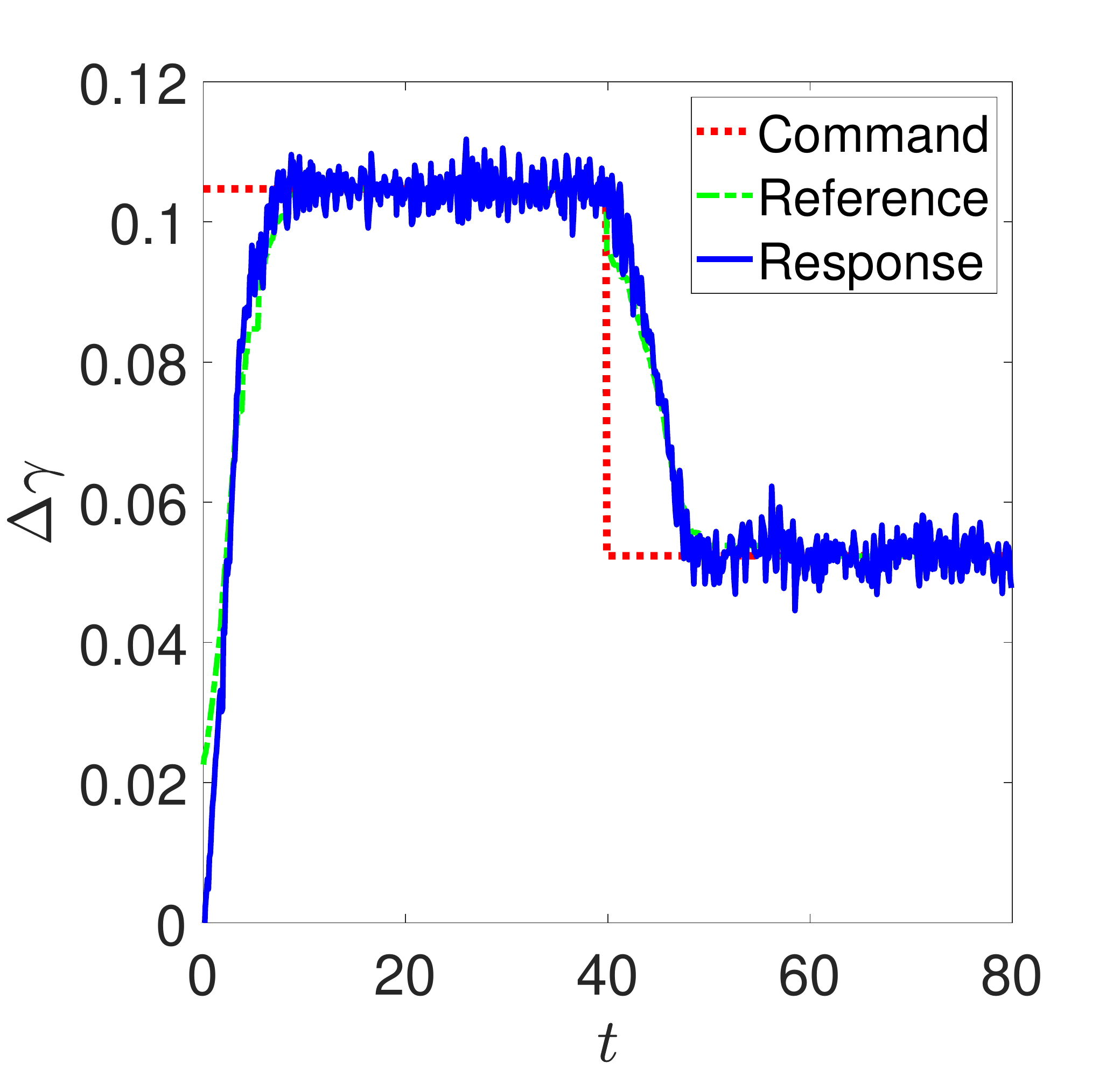,height=.24\textwidth}}
\put(  120,  0){\epsfig{file=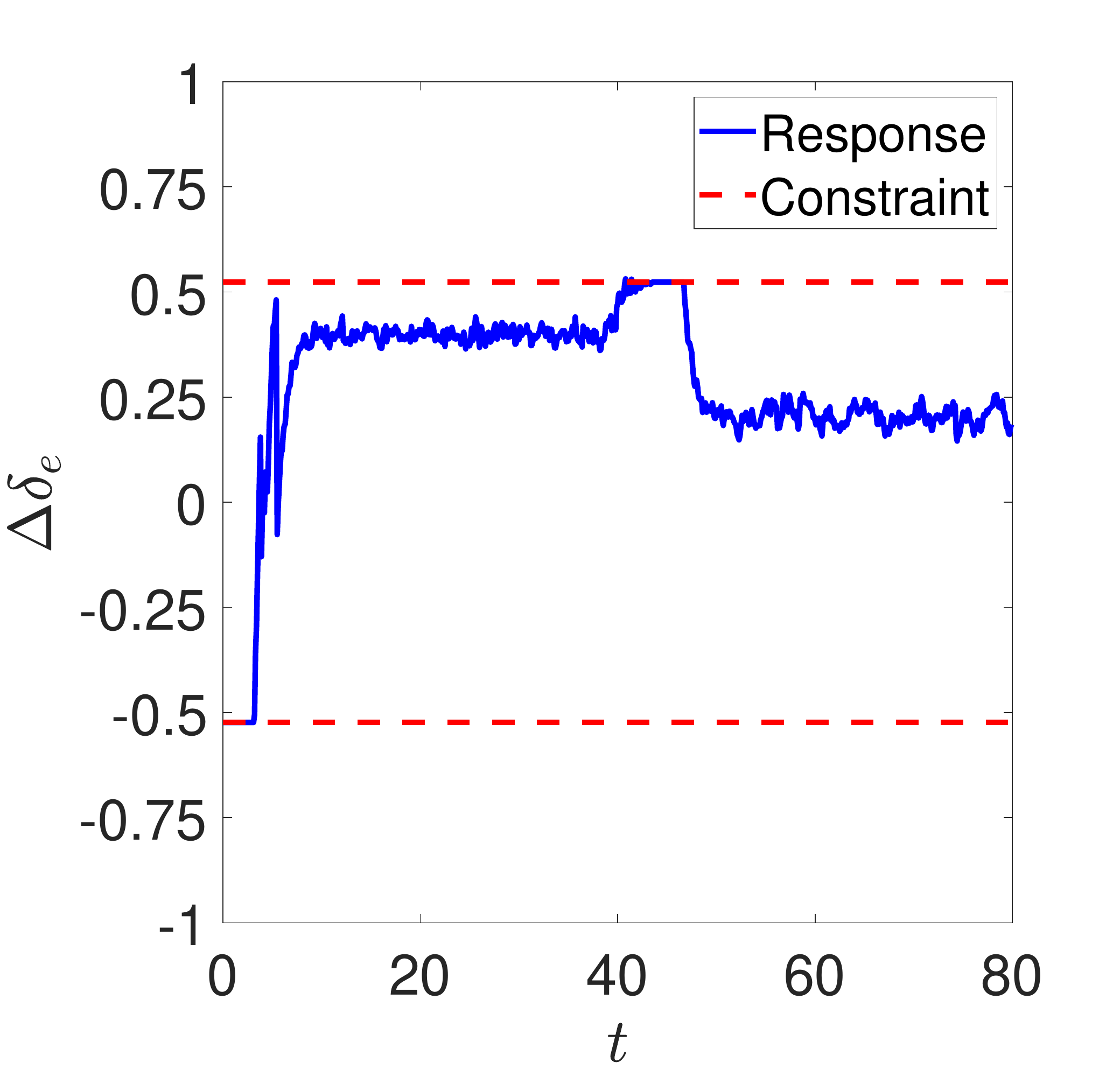,height=.24\textwidth}}
\small
\put( 100, 116){(a)}
\put( 220, 116){(b)}
\normalsize

\end{picture}
\end{center}
      \caption{Longitudinal flight control: Simulated trajectory of the closed-loop system augmented with CSRG Algorithm 1.}
      \label{fig:lon_2}
\end{figure}

\begin{figure}[ht!]
\begin{center}
\begin{picture}(240.0, 126.0)
\put(  0,  0){\epsfig{file=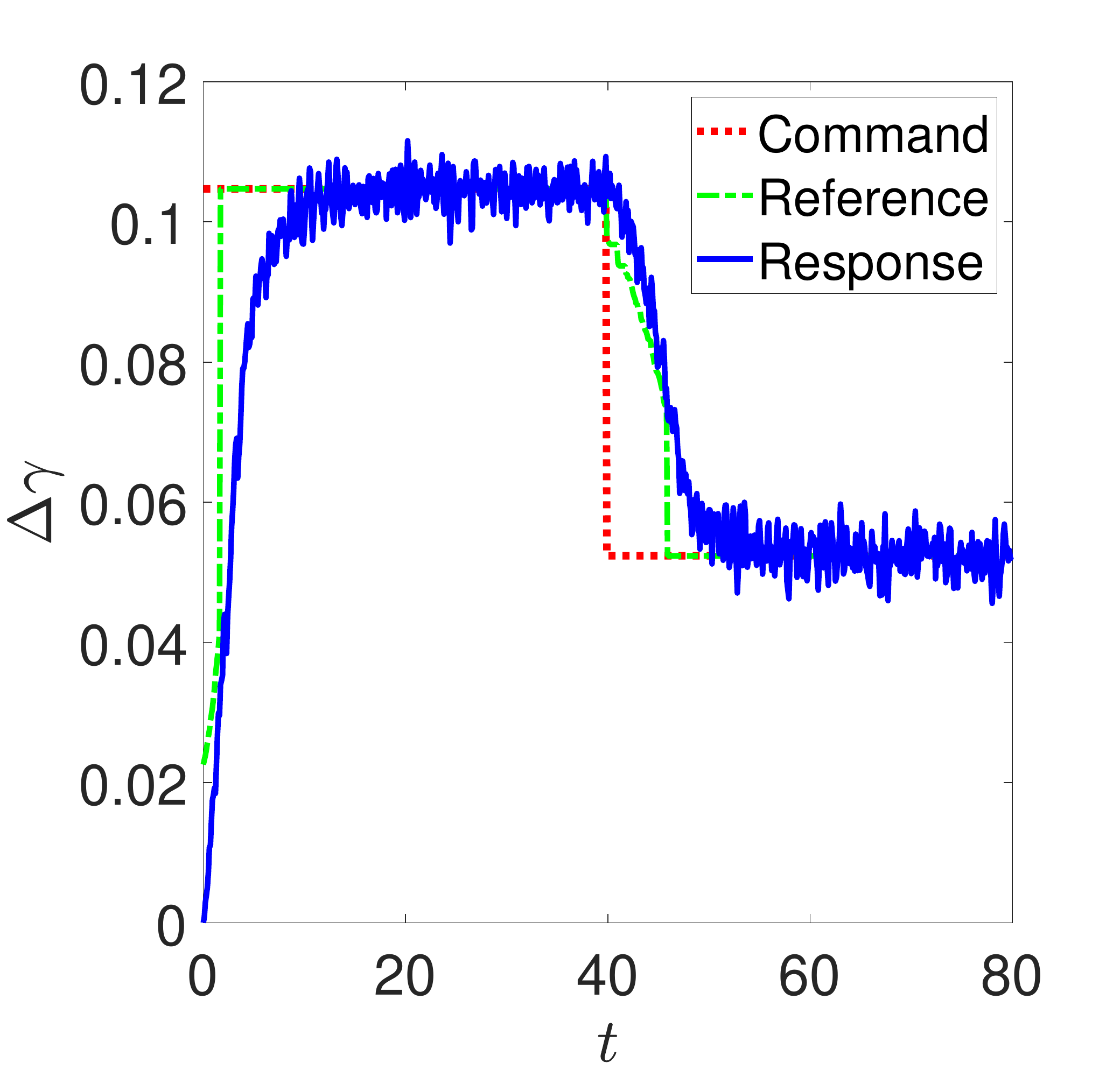,height=.24\textwidth}}
\put(  120,  0){\epsfig{file=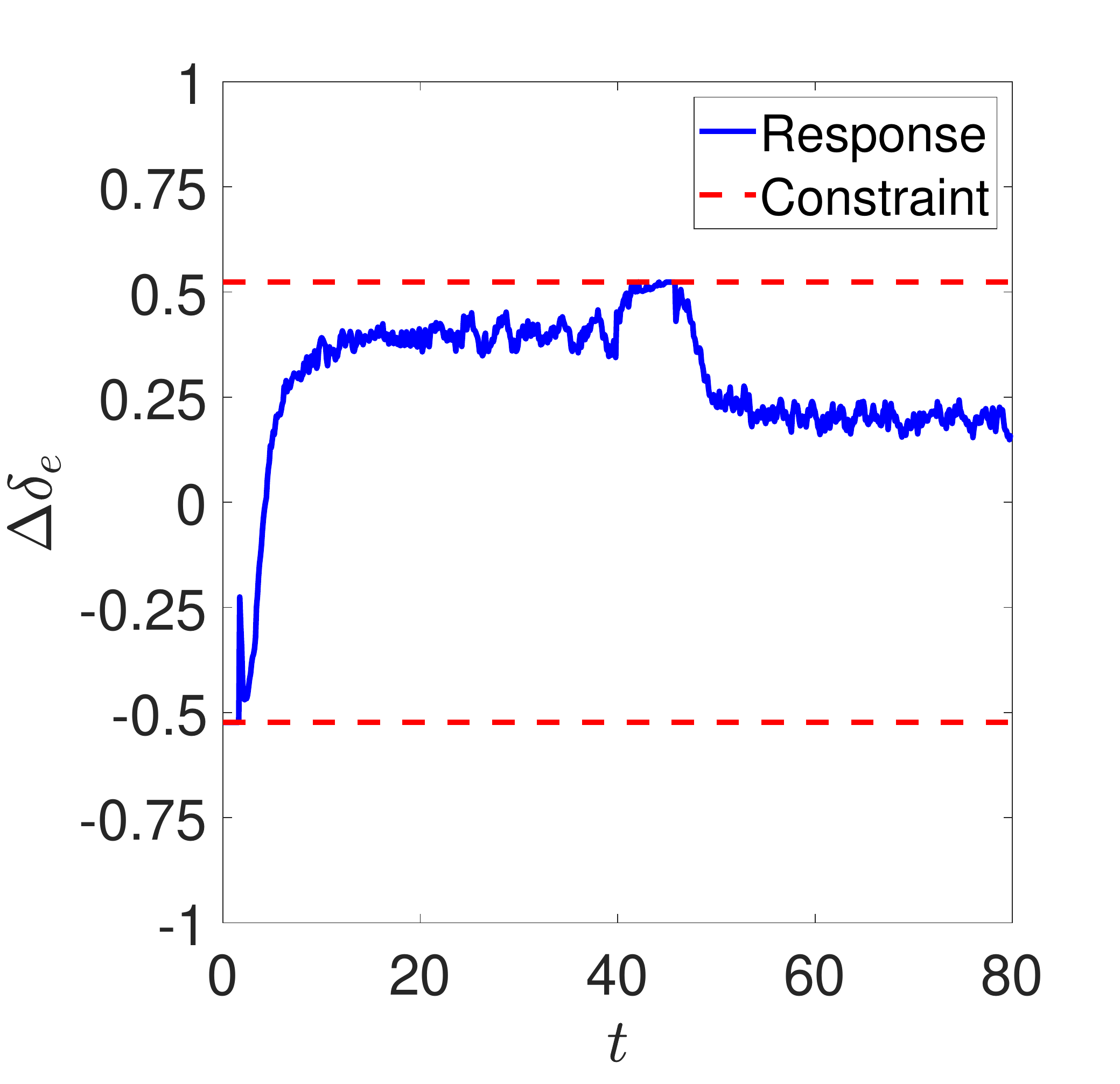,height=.24\textwidth}}
\small
\put( 100, 116){(a)}
\put( 220, 116){(b)}
\normalsize

\end{picture}
\end{center}
      \caption{Longitudinal flight control: Simulated trajectory of the closed-loop system augmented with CSRG Algorithm 2.}
      \label{fig:lon_3}
\end{figure}

We consider both the applications of CSRG Algorithms~1 and~2 and compare them. The responses of the closed-loop system starting from the same initial condition to track the same commanded flight path angle profile as before when augmented with CSRG Algorithms~1 and 2 are illustrated in Figs.~\ref{fig:lon_2} and \ref{fig:lon_3}, respectively. Recall that the CSRG replaces the original command $\gamma_{\text{com}}(t)$ in \eqref{equ:longitudinal_control_2} with a modified reference $v(t)$ to enforce constraints. The trajectory of $v(t)$ is shown by the green dashed-dotted curve in panel~(a) of Figs.~\ref{fig:lon_2} and \ref{fig:lon_3}. It can be observed that, in both cases, $v(t)$ deviates from the command $\gamma_{\text{com}}(t)$ when $\gamma_{\text{com}}(t)$ has step changes and converges to $\gamma_{\text{com}}(t)$ after a short-period transient response. From panel~(b) of Figs.~\ref{fig:lon_2} and \ref{fig:lon_3} we can observe that with CSRG, the significant violations of the constraints $-\frac{\pi}{6} \le \Delta \delta_e(t) \le \frac{\pi}{6}$ in Fig.~\ref{fig:lon_1}(b) have been avoided. Indeed, if we look at panel~(b) of Figs.~\ref{fig:lon_2} and \ref{fig:lon_3} more closely, we can find a few slight constraint violations over $40 \sim 45$ s. Such slight constraint violations are due to our probabilistic enforcement of the constraints. We also note that the other constraints of \eqref{equ:longitudinal_constraints} are all strictly satisfied in both cases.

Comparing panel~(a) of Figs.~\ref{fig:lon_2} and \ref{fig:lon_3} we can observe that, unlike the gradual convergence of $v(t)$ to $\gamma_{\text{com}}(t)$ in Fig.~\ref{fig:lon_2}(a), the modified reference $v(t)$ converges to the command $\gamma_{\text{com}}(t)$ abruptly through a jump at about $2$~s and through another jump at about $45$ s in Fig.~\ref{fig:lon_3}(a). This is a result of the fact that CSRG Algorithm~2 is designed for prioritizing the convergence of $v(t)$. Meanwhile, it can also be observed that the state responses resulting from the two algorithms, shown by the blue solid curves, have similar speeds in this example. Interestingly, in Fig.~\ref{fig:lon_2}(a) we can observe a few places where the state response is ahead of the reference response. This is partly attributed to the integrator state resets by CSRG, and is also related to effects of the stochastic disturbances \eqref{equ:longitudinal_disturbance}.

\subsection{Constrained lateral flight control}

The aircraft lateral dynamics are represented by the following continuous-time linear model, 
\vspace{-4mm}\small 
\begin{align}\label{equ:lateral_sys}
& \begin{bmatrix} \Delta \dot{\beta} \\ \Delta \dot{p} \\ \Delta \dot{r} \\ \Delta \dot{\phi} \end{bmatrix} = \underbrace{\begin{bmatrix} -0.5229 & 0.0861 & -0.9852 & 0.2374 \\
-90.5885 & -6.2736 & 2.0861 & 0 \\
29.1873 & -0.4833 & -1.4043 & 0 \\
0 & 1 & 0.0857 & 0 \end{bmatrix}}_{= A^{\text{lat}}} \begin{bmatrix} \Delta \beta \\ \Delta p \\ \Delta r \\ \Delta \phi \end{bmatrix} \nonumber \\
& + \underbrace{\begin{bmatrix} -0.0002 & 0.0031 \\
-0.9174 & 0.2321 \\
-0.0523 & -0.4436 \\
0 & 0 \end{bmatrix}}_{= B_u^{\text{lat}}} \begin{bmatrix} \Delta \delta_a \\ \Delta \delta_r \end{bmatrix} - \underbrace{\begin{bmatrix} -0.5229 \\
-90.5885 \\
29.1873 \\
0 & \end{bmatrix}}_{= B_w^{\text{lat}}} w_y,
\end{align}
\normalsize
where the states $x_p^{\text{lat}} = [\Delta \beta, \Delta p, \Delta r, \Delta \phi]^{\top}$ represent the deviations in sideslip angle (rad), roll rate (rad/s), yaw rate (rad/s) and roll angle (rad), respectively, the control inputs $u^{\text{lat}} = [\Delta \delta_a, \Delta \delta_r]^{\top}$ represent the deviations in aileron angle (rad) and rudder angle (rad), and the disturbance input $w^{\text{lat}} = w_y$ represents effects of model mismatch, atmospheric disturbances/turbulence, etc. in the lateral direction.

Firstly, as in the longitudinal case, the continuous-time model \eqref{equ:lateral_sys} is converted to a discrete-time model using zero-order hold on the inputs with a sampling period of $\Delta T = 0.1$ s. Then, a nominal controller in the form of \eqref{equ:system_controller} with
\begin{align}\label{equ:lateral_control}
K_p^{\text{lat}} &= \begin{bmatrix} -1.4874 & 0.3021 & 0.8549 & 2.1801 \\
   -0.4431 & 0.2363 & 1.2214 & 2.1289 \end{bmatrix}, \nonumber \\
K_u^{\text{lat}} &=  \begin{bmatrix} 0.0680 \\ 0.0684 \end{bmatrix}, \quad\quad B_v^{\text{lat}} = \begin{bmatrix} 0 \\ 0 \end{bmatrix}, \\
A_p^{\text{lat}} &= \big[\, 0\;\; 0\;\; 0\;\; 1\, \big],\quad\quad A_u^{\text{lat}} = 1, \quad\quad D_v^{\text{lat}} = -1, \nonumber
\end{align}
is used to track commanded roll angles $r(t) = \Delta \phi_{\text{com}}(t)$ \citep{mcdonough2015integrator}, where the gains $K^{\text{lat}} = [K_p^{\text{lat}}, K_u^{\text{lat}}]$ are obtained by solving a discrete-time LQR problem with $\bar{A}_{\text{open}}^{\text{lat}} = \begin{bmatrix} A^{\text{lat}} & 0 \\ A_p^{\text{lat}} & A_u^{\text{lat}} \end{bmatrix}$, $\bar{B}_u^{\text{lat}} = \begin{bmatrix} B_u^{\text{lat}} \\ 0 \end{bmatrix}$ and $Q^{\text{lat}} = \text{diag}(100, 100, 100, 100, 1)$, $R^{\text{lat}} = \text{diag}(100, 100)$.

We consider the following set of constraints,
\begin{align}\label{equ:lateral_constraints}
& -\frac{\pi}{12} \le \Delta \beta \le \frac{\pi}{12}, &  -\frac{\pi}{12} \le \Delta p \le \frac{\pi}{12}, & \nonumber \\
& -\frac{\pi}{12} \le \Delta r \le \frac{\pi}{12}, & -\frac{\pi}{3} \le \Delta \phi \le \frac{\pi}{3}, & \\
& -\frac{\pi}{6} \le \Delta \delta_a \le \frac{\pi}{6}, & -\frac{\pi}{6} \le \Delta \delta_r \le \frac{\pi}{6}, \nonumber &
\end{align}
and we model the disturbance $w^{\text{lat}} = w_y$ as a Gaussian noise satisfying Assumption~2 with
\begin{equation}\label{equ:lateral_disturbance}
\{w^{\text{lat}}(t)\}_{t \in \mathbb{Z}_{\ge 0}} \sim \mathcal{N} \left(0, 10^{-5}\right).
\end{equation}

In this example, we enforce the constraints \eqref{equ:lateral_constraints} jointly as a chance constraint in the form of \eqref{equ:joint_constraint}, where the required confidence level $\beta$ is chosen as $0.98$. It can be determined from \eqref{equ:lateral_constraints} that we have $n_y = 6$ outputs and $n_g = 12$ constraints. In this case, referring to Proposition~5 or Fig.~\ref{fig:Gamma}, we know that the risk allocation approach has a lower degree of conservativeness than the confidence ellipsoid approach for treating the formulated joint chance constraint. Therefore, we choose to use the risk allocation approach to formulate our CSRG algorithm. In particular, according to \eqref{equ:risk_allocation_2}, we set $\beta_i = \frac{0.98 + (12 - 1)}{12} = 0.9983$  for all $i = 1,...,12$. The parameters $P$, $R$ and $\delta$ for the CSRG online optimization problem are chosen to be the same as in the longitudinal case.

\begin{figure}[ht!]
\begin{center}
\begin{picture}(240.0, 368.0)
\put(  0,  240){\epsfig{file=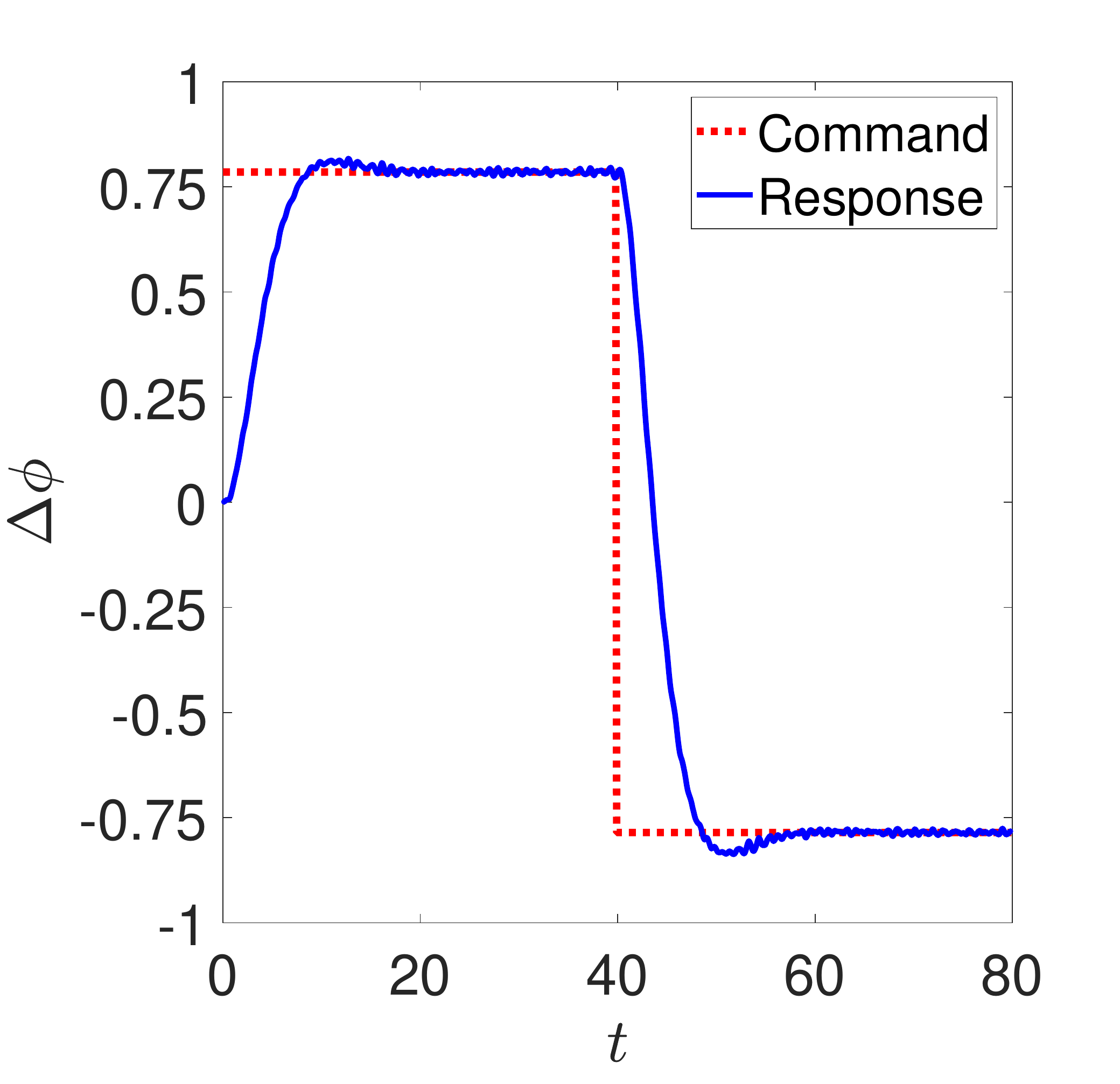,height=.24\textwidth}}
\put(  0,  120){\epsfig{file=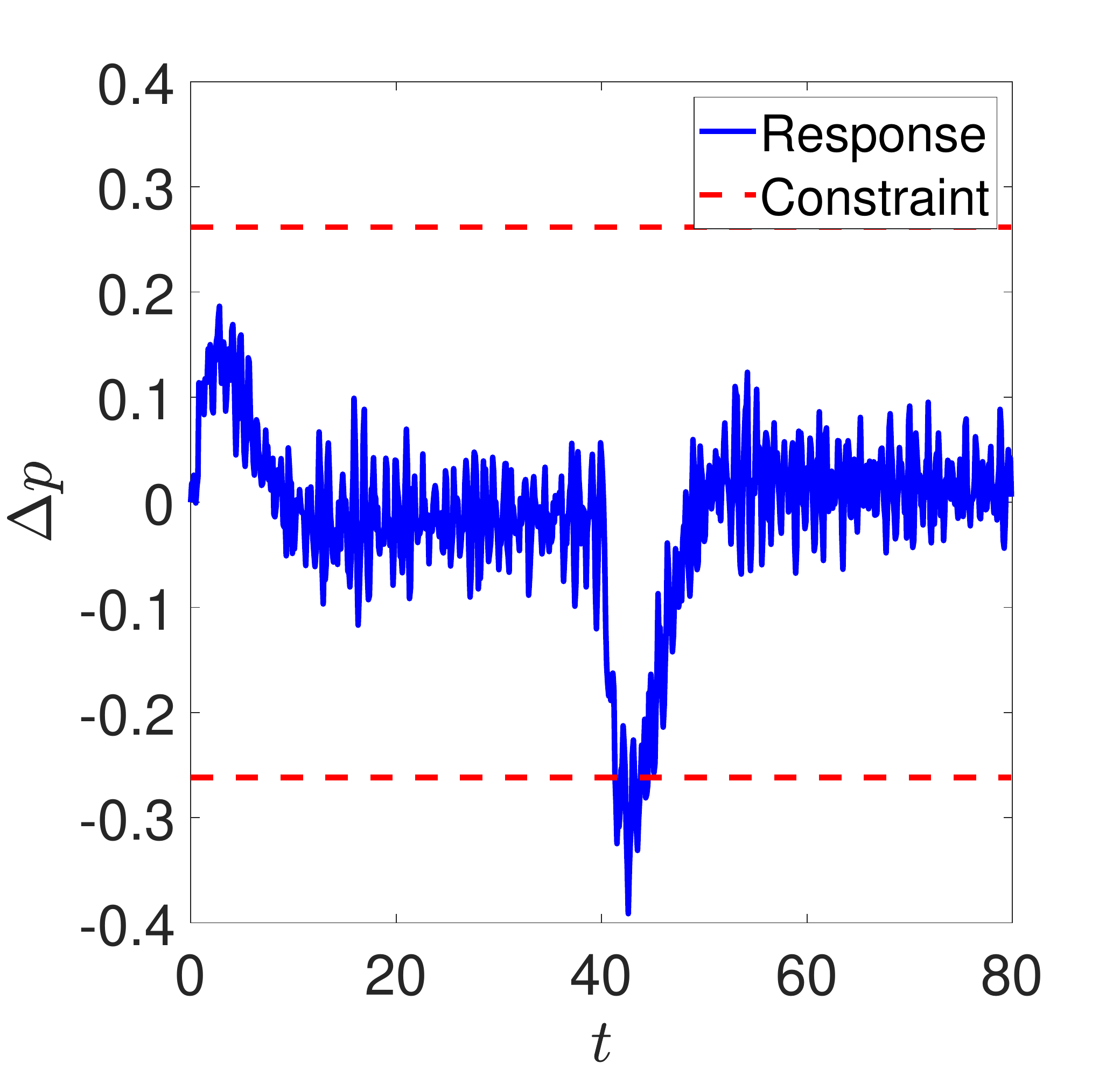,height=.24\textwidth}}
\put(  0,  0){\epsfig{file=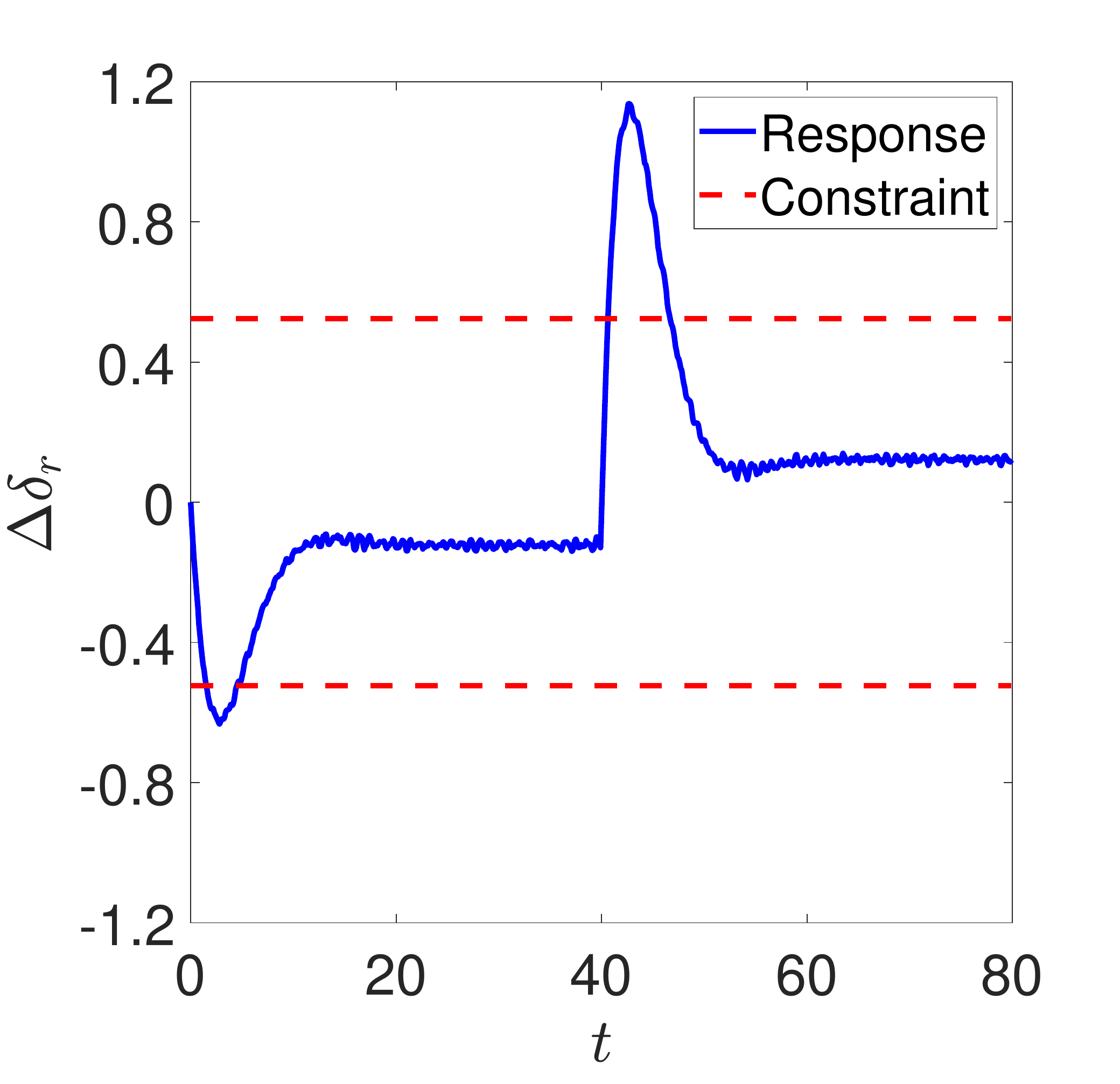,height=.24\textwidth}}
\put(  120,  240){\epsfig{file=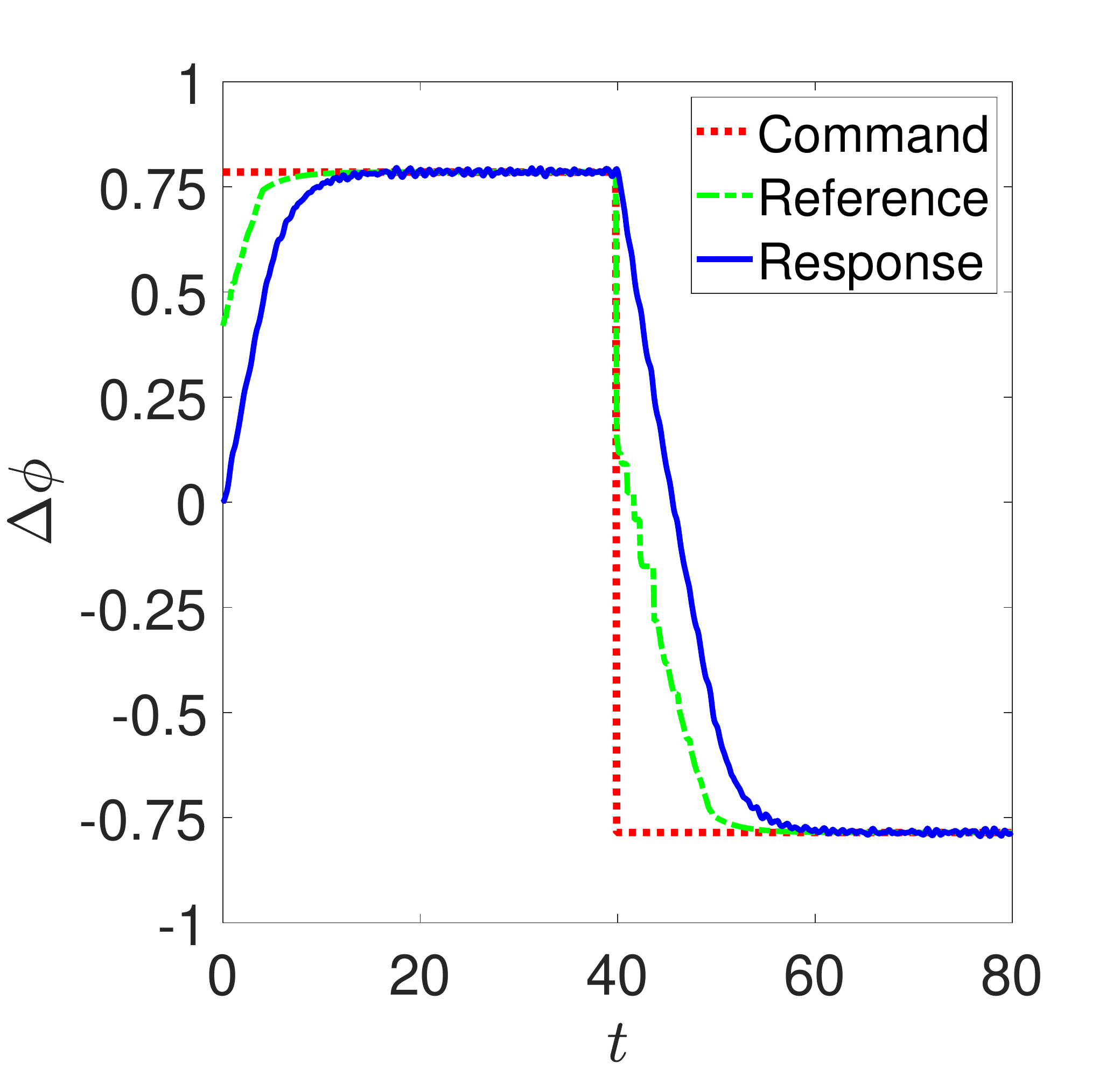,height=.24\textwidth}}
\put(  120,  120){\epsfig{file=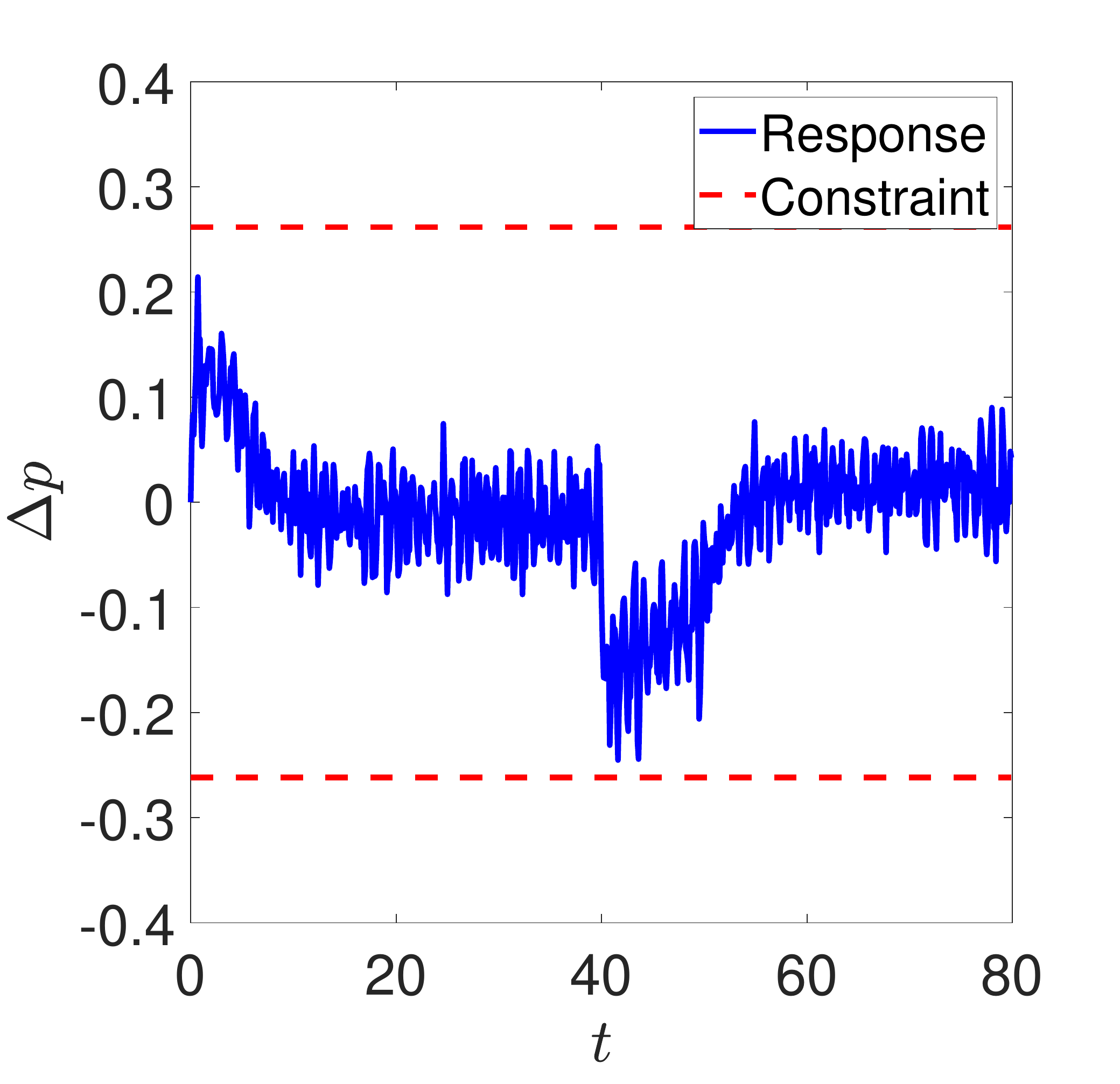,height=.24\textwidth}}
\put(  120,  0){\epsfig{file=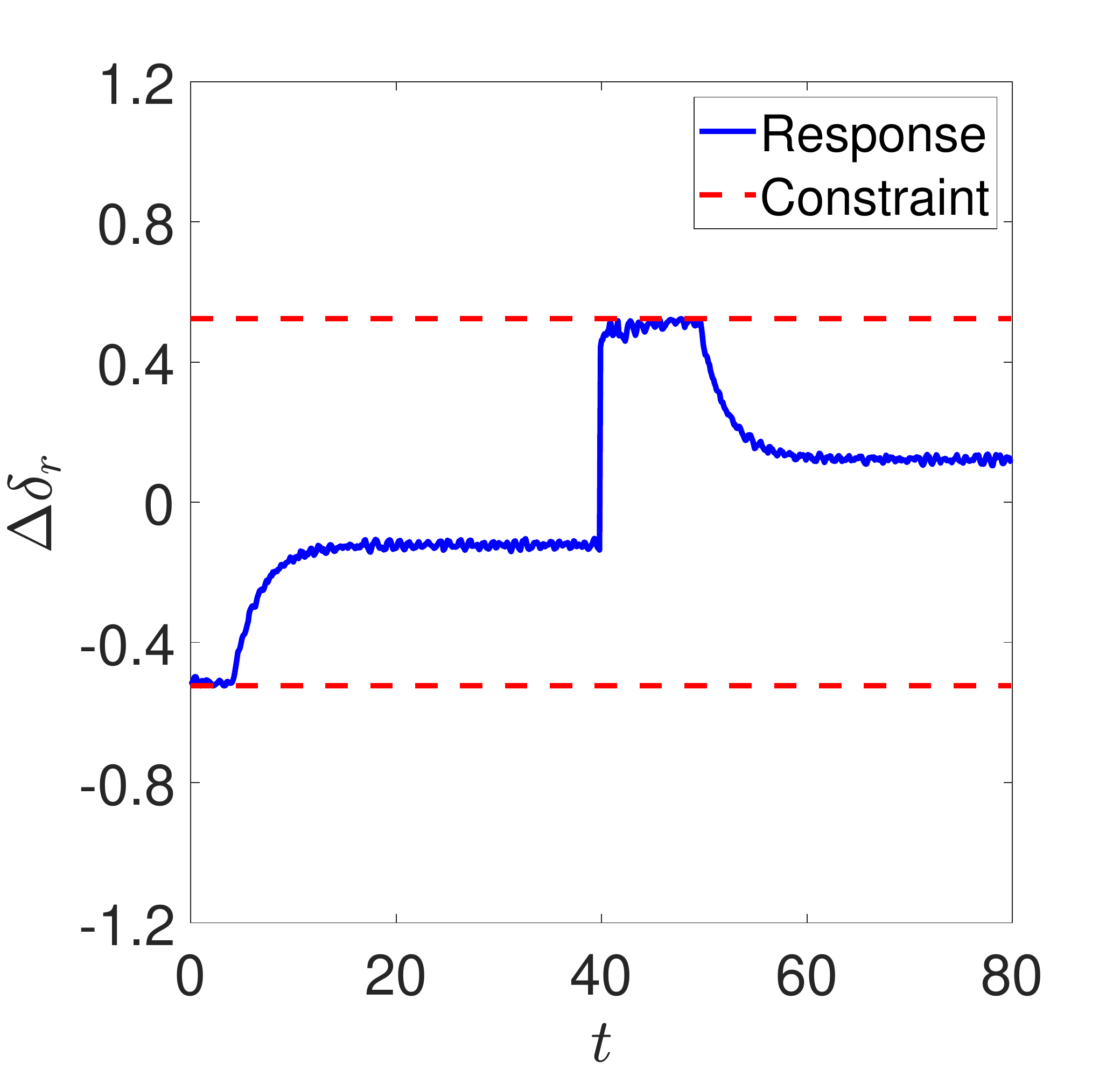,height=.24\textwidth}}
\small
\put( 94, 356){(a-1)}
\put( 214, 356){(b-1)}
\put( 94, 236){(a-2)}
\put( 214, 236){(b-2)}
\put( 94, 116){(a-3)}
\put( 214, 116){(b-3)}
\normalsize

\end{picture}
\end{center}
      \caption{Lateral flight control: Simulated trajectories of (a) the nominal closed-loop system and (b) the closed-loop system augmented with CSRG Algorithm 1.}
      \label{fig:lat}
\end{figure}
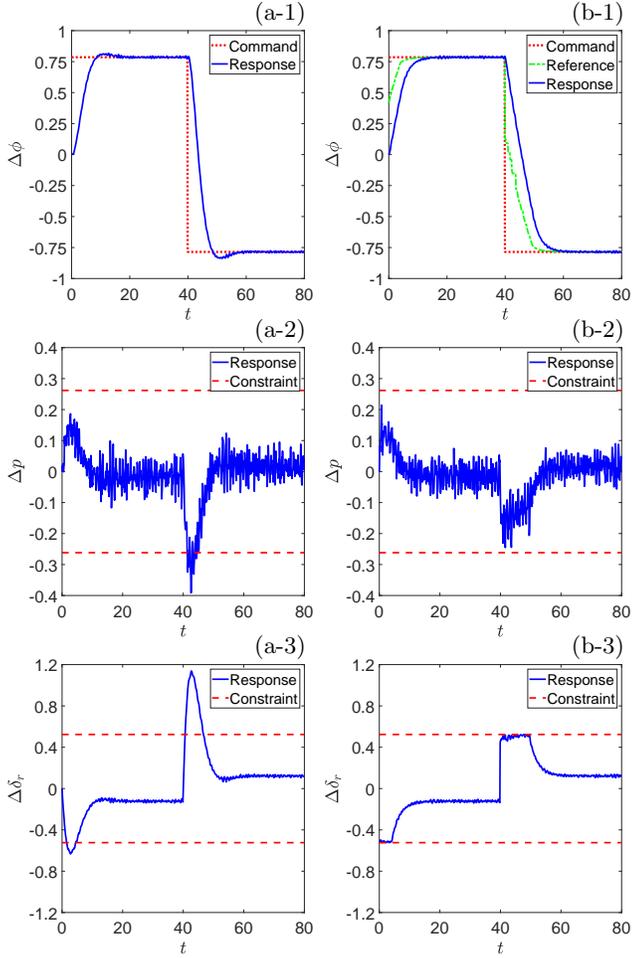

The state response of the nominal closed-loop system, i.e., without CSRG, and the reference and state responses of the closed-loop system augmented with CSRG Algorithm~1 starting from the initial condition $x_p^{\text{lat}}(0) = [0,0,0,0]^{\top}$ and $x_u^{\text{lat}}(0) = 0$ to track the commanded roll angle profile $\Delta \phi_{\text{com}}(t)$ shown by the red dotted curve in Fig.~\ref{fig:lat}(a-1) are illustrated in Fig.~\ref{fig:lat}(a) and (b), respectively. It can be observed from Fig.~\ref{fig:lat} that without CSRG, the constraints on $\Delta p$ and on $\Delta \delta_r$ are violated; with CSRG, these constraint violations are avoided. We note that the other constraints of \eqref{equ:lateral_constraints} are also enforced when CSRG is used.

\subsection{Comparisons}

To compare the constrained domain of attraction with both controller state and reference modifications and that with only reference modification (i.e., without controller state modification), we define the following set,
\begin{equation}
    \tilde{\mathcal{O}}_{\infty}^{x_u(0) = 0} = \tilde{\mathcal{O}}_{\infty} \cap \big\{(x_p,x_u,v):\, x_u = 0\big\}.
\end{equation}
The set $\tilde{\mathcal{O}}_{\infty}^{x_u(0) = 0}$ represents the set of chance-constraint admissible initial plant state and constant reference input pairs $\left(x_p(0),v\right)$ when the initial controller state $x_u(0)$ is zero. Then, we consider the projections of $\tilde{\mathcal{O}}_{\infty}$ and $\tilde{\mathcal{O}}_{\infty}^{x_u(0) = 0}$ onto the $x_p$-space, denoted as ${\rm proj}_{x_p}(\tilde{\mathcal{O}}_{\infty})$ and ${\rm proj}_{x_p}(\tilde{\mathcal{O}}_{\infty}^{x_u(0) = 0})$, respectively. 

According to the definition of $\tilde{\mathcal{O}}_{\infty}$, ${\rm proj}_{x_p}(\tilde{\mathcal{O}}_{\infty})$ represents the set of plant states at which there exist controller state and reference pairs that guarantee their corresponding future system trajectories satisfying the chance constraints. Referring to Propositions~3 and 4, the reference $v(t)$ and state $\bar{x}(t)$ responses in closed-loop operation of the system augmented with CSRG starting from these plant states are guaranteed to converge to $r$ in finite time almost surely and to $\bar{x}^*(r)$ in mean square for constant, steady-state constraint-admissible command $r$. Therefore, ${\rm proj}_{x_p}(\tilde{\mathcal{O}}_{\infty})$ is referred to as the constrained domain of attraction of CSRG. In contrast, ${\rm proj}_{x_p}(\tilde{\mathcal{O}}_{\infty}^{x_u(0) = 0})$ represents the set of plant states at which the chance constraints can be enforced by properly choosing the reference value if the current controller state is zero and cannot be adjusted.

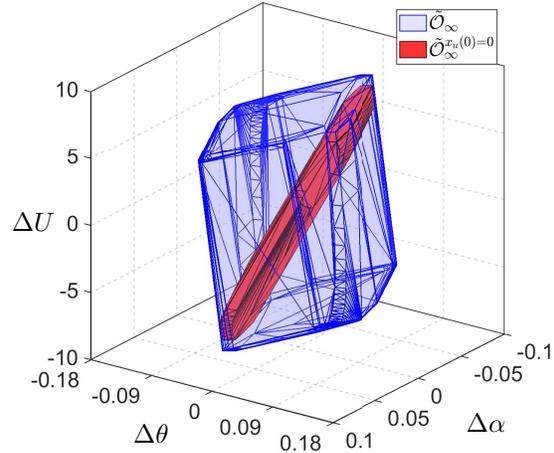
\begin{figure}[h!]
\begin{center}
\begin{picture}(210.0, 183.5)
\put(  0,  -4){\epsfig{file=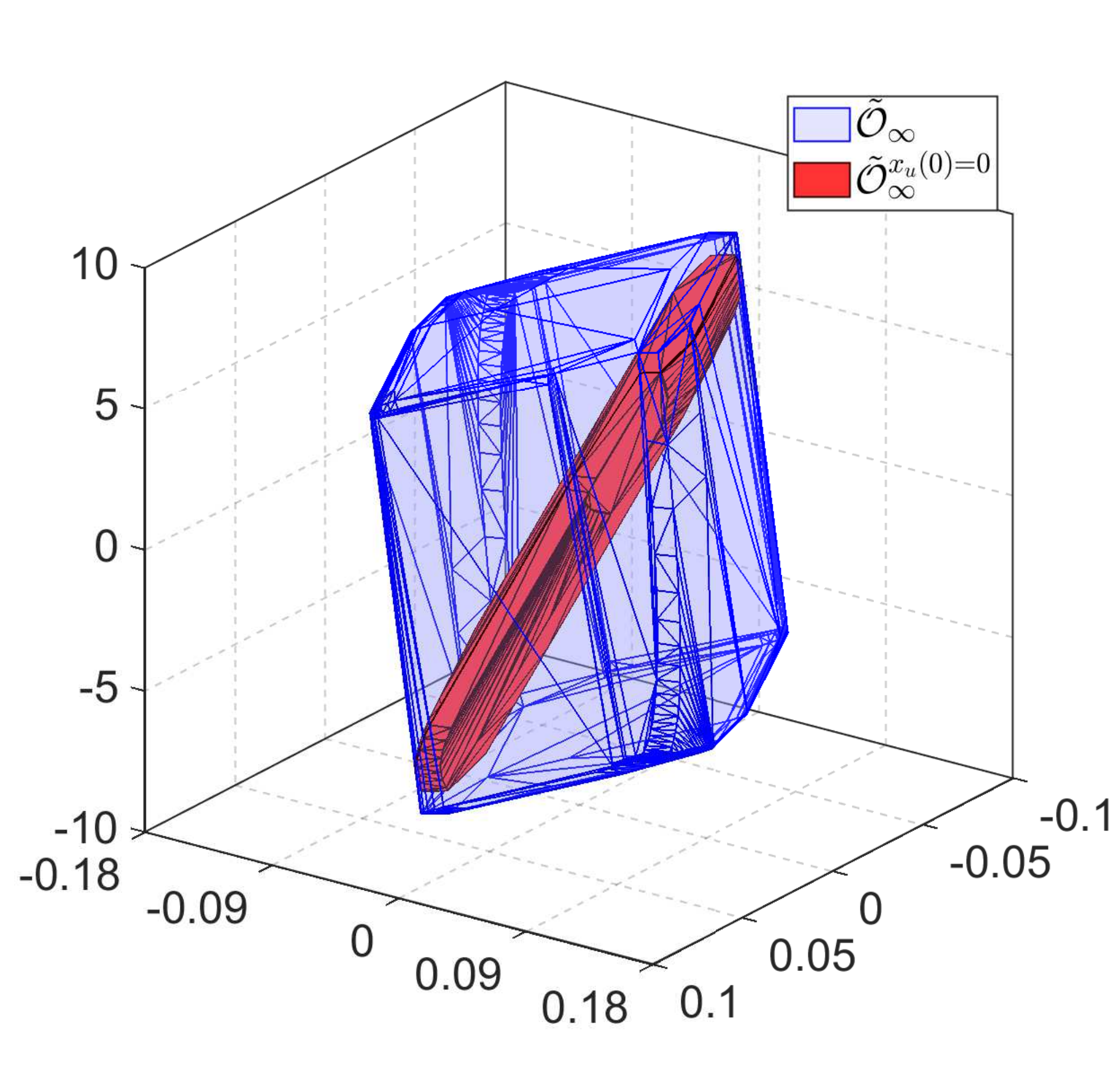,height=2.7in}}  
\put(  42,  10){$\Delta \theta$}
\put(  168,  14){$\Delta \alpha$}
\put(  -4,  90){$\Delta U$}
\end{picture}
\end{center}
      \caption{Longitudinal flight control: Projections of $\tilde{\mathcal{O}}_{\infty}$ versus $\tilde{\mathcal{O}}_{\infty}^{x_u(0) = 0}$ onto the $(\Delta U,\Delta \alpha,\Delta \theta)$-space.}
      \label{fig:O_inf_lon}
\end{figure}

\begin{figure}[h!]
\begin{center}
\begin{picture}(200.0, 180.0)
\put(  0,  0){\epsfig{file=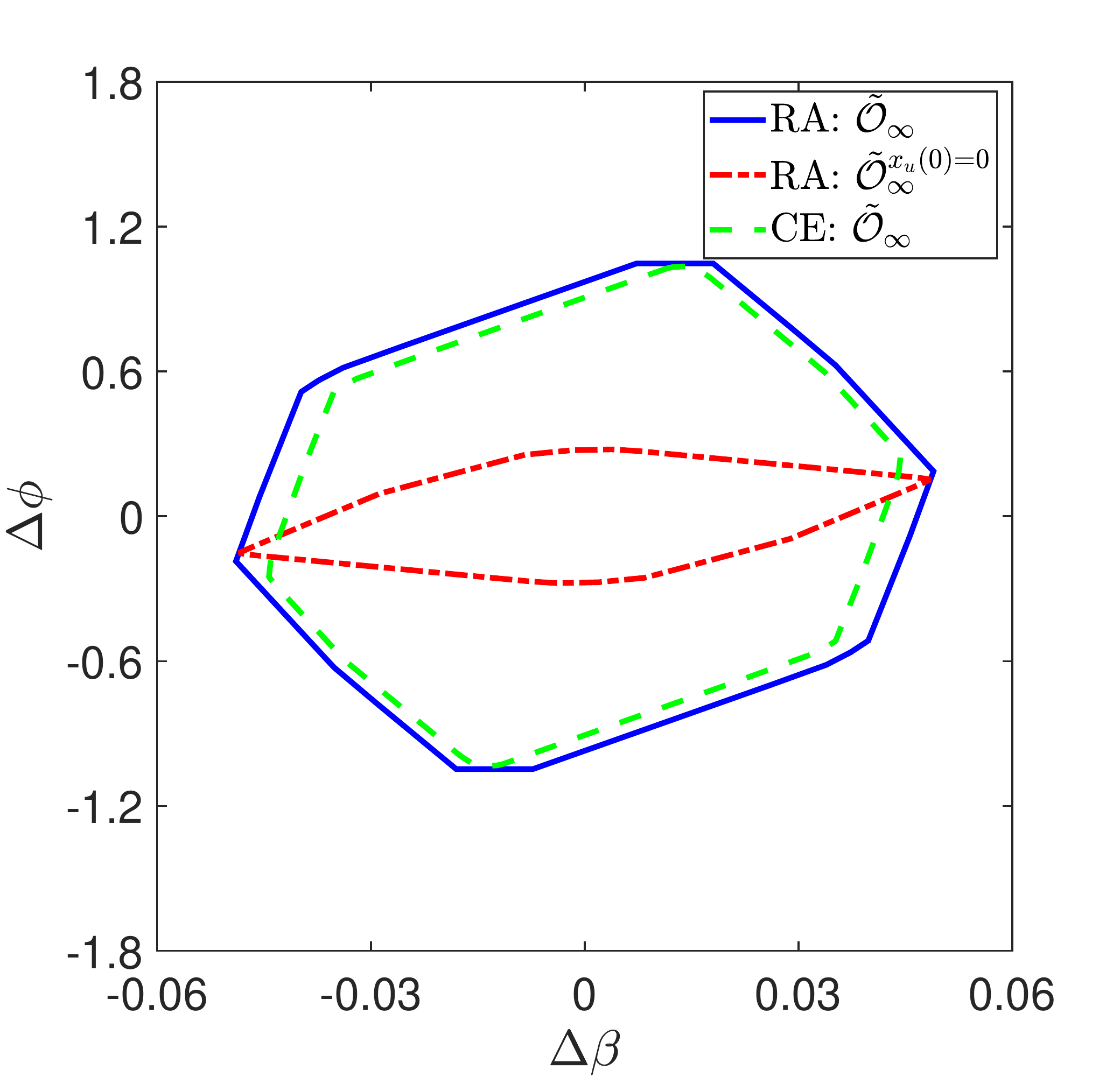,height=2.5in}}  
\end{picture}
\end{center}
      \caption{Lateral flight control: Projections of $\tilde{\mathcal{O}}_{\infty}$, $\tilde{\mathcal{O}}_{\infty}^{x_u(0) = 0}$ with the risk allocation (RA) approach, and $\tilde{\mathcal{O}}_{\infty}$ with the confidence ellipsoid (CE) approach onto the $(\Delta \beta,\Delta \phi)$-plane.}
      \label{fig:O_inf_lat}
\end{figure}

The comparison between ${\rm proj}_{x_p}(\tilde{\mathcal{O}}_{\infty})$ and ${\rm proj}_{x_p}(\tilde{\mathcal{O}}_{\infty}^{x_u(0) = 0})$ for the longitudinal flight control example is illustrated in Fig.~\ref{fig:O_inf_lon}, where the blue and red 3D polygons show, respectively, the projections of $\tilde{\mathcal{O}}_{\infty}$ and $\tilde{\mathcal{O}}_{\infty}^{x_u(0) = 0}$ onto the $(\Delta U,\Delta \alpha,\Delta \theta)$-space. Similarly, ${\rm proj}_{x_p}(\tilde{\mathcal{O}}_{\infty})$ and ${\rm proj}_{x_p}(\tilde{\mathcal{O}}_{\infty}^{x_u(0) = 0})$ for the lateral flight control example are illustrated by the blue solid and red dash-dotted 2D polygons in Fig.~\ref{fig:O_inf_lat}. It can be observed from Figs.~\ref{fig:O_inf_lon} and \ref{fig:O_inf_lat} that ${\rm proj}_{x_p}(\tilde{\mathcal{O}}_{\infty})$ is much larger than ${\rm proj}_{x_p}(\tilde{\mathcal{O}}_{\infty}^{x_u(0) = 0})$ in both cases. This demonstrates the fact that, by admitting both controller state and reference modifications, our proposed CSRG scheme can have a considerably larger constrained domain of attraction compared to admitting only reference modification as in conventional RG schemes.

In Fig.~\ref{fig:O_inf_lat}, we also show the projection onto the $(\Delta \beta,\Delta \phi)$-plane of the $\tilde{\mathcal{O}}_{\infty}$ set corresponding to the confidence ellipsoid approach to treating the formulated joint chance constraint in the lateral flight control example. It can be observed that the green dashed polygon, which corresponds to the confidence ellipsoid approach, is strictly contained in the blue solid polygon, which corresponds to the risk allocation approach. The conclusion that the risk allocation approach is less conservative than the confidence ellipsoid approach in this example is consistent with Proposition~5 and Fig.~\ref{fig:Gamma}. Specifically, in this example we have $(n_y,n_g,\beta) = (6,12,0.98)$ and $\Gamma(n_y,n_g,\beta) = \Gamma(6,12,0.98) < 0$.

\section{Concluding Remarks}\label{sec:6}

In this paper, we introduced the chance-constrained controller state and reference governor (CSRG), as an add-on scheme for closed-loop systems with dynamic controllers that are subject to stochastic disturbances and pointwise-in-time constraints. We showed that this chance-constrained CSRG guarantees closed-loop chance-constraint satisfaction, almost-sure finite-time convergence of the modified reference to constant, steady-state constraint-admissible command, and mean-square stability of the commanded state set-point. We also extended CSRG formulation from treating individual chance constraints to treating joint chance constraints, and we have developed guidelines for when risk allocation is advantageous over the confidence ellipsoid approach in treating such joint constraints. Finally, we illustrated CSRG application using constrained aircraft flight control examples.

\section*{Acknowledgments}

This research was supported by the National Science Foundation awards ECCS~1931738 and CMMI~1904394.

\bibliographystyle{ifacconf-harvard}
\bibliography{ref}

\end{document}